\documentclass[twocolumn]{autart}    
\usepackage{psfrag} 
\usepackage{graphicx}          
\usepackage{float}
\usepackage{amsmath}
\usepackage{algorithm}
\usepackage{algorithmicx}
\usepackage{algpseudocode}
\usepackage{subcaption}
\usepackage{amsmath,amssymb}
\usepackage{amsfonts}
\usepackage{bm}
\usepackage[round]{natbib}
\usepackage{hyperref}
\hypersetup{
    colorlinks=true,
    linkcolor=blue,
    citecolor=cyan,
    urlcolor=blue
}
\usepackage[utf8]{inputenc}
\usepackage{babel}
\usepackage{amsmath}
 \DeclareUnicodeCharacter{221E}{$\infty$}

\begin{document}

\begin{frontmatter}

\title{State Anti-windup: A New Methodology for Tackling State Constraints at Both Synthesis and Implementation Levels}

\thanks[footnoteinfo]{This paper was not presented at any IFAC 
meeting. Corresponding authorA.~H.~Abolmasoumi.}
\author[ECN]{Amir H. Abolmasoumi}\ead{amir.abolmasoumi@ec-nantes.fr},    
\author[ECN]{Bogdan Marinescu}\ead{bogdan.marinescu@ec-nantes.fr},               

\address[ECN]{Ecole Centrale de Nantes-LS2N, Nantes, France}  
          

\begin{abstract}

\noindent The anti-windup compensation typically addresses strict control limitations in control systems. However, there is a clear need for an equivalent solution for the states/outputs of the system. This paper introduces a novel methodology for the state anti-windup compensator. Unlike state-constrained control methods, which often focus on incorporating soft constraints into the design or fail to react adequately to constraint violations in practical settings, the proposed methodology treats state constraints as implement-oriented ‘soft-hard’ constraints. This is achieved by integrating a saturation block within the structure of the safety compensator, referred to as the state anti-windup (SANTW) compensator. Similar to input anti-windup schemes, the SANTW design is separated from the nominal controller design. The problem is formulated as a disturbance rejection one to directly minimize the saturation. The paper develops two $H_{\infty}$ optimization frameworks using frequency-domain solutions and linear matrix inequalities. It then addresses constraints on both inputs and states, resulting in a unified Input-State Anti-windup (IS-ANTW) compensator synthesized using non-smooth $H_{\infty}$ optimization. This method also offers the flexibility of having a fixed-order compensator, crucial in many practical applications. Additionally, the study evaluates the proposed compensator's performance in managing current fluctuations from renewable energy sources during grid faults, demonstrating its effectiveness through detailed Electromagnetic Transient (EMT) simulations of grid-connected DC-AC converters.

\end{abstract}

\end{frontmatter}

\section{Introduction}
In real-world dynamic systems, constraints on inputs arise due to various reasons such as actuator limitations, protection measures, or safety devices. To address these strict constraints and to prevent issues caused by saturation, methods known as ‘\textit{anti-windup}’ have been developed. These methods adjust the nominal control within a control system to adhere to such ‘\textit{hard}’ constraints. By enforcing limits on the input signals, anti-windup methods effectively reduce the risk of control system instability and performance degradation. The literature on input anti-windup methods is extensive, spanning many years. For a comprehensive overview of both classic and modern anti-windup compensation methods, refer to seminal works such as \cite{galeani2009tutorial, tarbouriech2011stability, zaccarian2011modern, kothare1994unified, mulder2001multivariable}. Additionally, recent advancements in this field can be found in \cite{moreno2023limited, lai2024performance}.

While there are hard constraints on the inputs of a control system, no equivalent constraints exist for state/output variables. Apart from safety protective shut-down tools, there are no definitive solutions to limit states or outputs. Consequently, an equivalent anti-windup concept to address the state constraints is currently unavailable, despite its practical necessity. This need is evident in several applications, such as grid-connected DC-AC converters, which are critical for integrating renewable energy resources into the grid. In these systems, enforcing strict limitations on states—such as currents—is mandatory for practical implementation. Implementing such constraints would significantly enhance fault-ride-through capability and improve resilience in over-current situations, thereby ensuring the stability and reliability of grid-connected renewable energy resources (see, for example, \cite{zhong2016current, taul2019current} and the references therein). Indeed, state constraints are usually addressed in a ‘\textit{soft}’  manner that is addressed during control synthesis and not on the implementation level. However, to create an equivalent concept of anti-windup for the states and outputs of a system, a novel methodology is required to bring it closer to the level of practical implementation. As mentioned, unlike traditional anti-windup, where input bound violations are managed by physical bounding devices integrated into the hardware, there is no such hardware limitations for state constraint violations. Instead, incorporating saturation checking explicitly within the compensator is a way to maintain the analogy with the anti-windup concept.

As discussed, efforts to constrain states in dynamic systems primarily focus on soft constraints, addressed during control synthesis and not on the implementation level. These constraints, defining permissible sets for system variables, are often integrated into optimal control formulations like Model Predictive Control (MPC) \cite{Rawlings2009, keusch2023long}, Linear Quadratic Regulator (LQR) methods \cite{Garcia1989, Bemporad2002} and more general Mixed-Integer Quadratic Programming (MIQP) \cite{Bemporad2002}. While these methods, along with more recent approaches by \cite{ames2016control} and \cite{wang2024adaptive}, aim to keep state trajectories within specified limits, they lack direct link to practical implementation, offering limited real-world protection. This constraint-integrated controller design may compromise nominal behavior in unconstrained scenarios, exacerbated by modeling uncertainties and external disturbances. Safety filters, a newer method in control systems, aim to add layers of protection \cite{fisac2018general, dhiman2021control, emam2022safe, wabersich2021probabilistic}, however, they are not yet ‘\textit{implementation-oriented}’, relying heavily on models and susceptible to uncertainties and disturbances. In fact, their divergence from hardware reality can result in a lack of \textit{'reaction'} when the ideal safety bounds are breached during real-time operation.

This paper introduces the concept of state anti-windup (SANTW) to address ‘\textit{soft-hard}’ constraints on states. By soft-hard constraints, we mean constraints integrated into the compensation structure using a saturation function applied to measured states/outputs. While the saturation function is not physically present in the real world, it ensures implementation-oriented design and responsiveness to violations of essential state/output constraints. Here, as the main contribution, the problem of state anti-windup based on soft-hard constraints is formulated as a general disturbance rejection problem, solvable using several methodologies. 

The corresponding optimization aimed at minimizing the impact of saturated states/outputs and nominal control on constraint violations, as well as the level of interference caused by the compensator. This novel formulation allows for a variety of solutions, ranging from nonlinear optimization frameworks to frequency-based $H_{\infty}$ and equivalent convex formulations. As another significant contribution, we provide a solution for the SANTW disturbance rejection problem using frequency-domain $H_{\infty}$ optimization and linear matrix inequalities (LMIs). As the ultimate yet pivotal contribution of the paper, to address hard limitations on the overall control input and to suggest a universal compensator capable of simultaneously handling both input saturation and state/output saturation, we introduce the so-called unified input-state anti-windup compensator (IS-ANTW) by developing two different approaches.

The performance of all proposed compensators is validated \textcolor{black}{first on a simple example of mathematical dynamic system. Next, results from a real industrial application in limiting over-current in a grid-connected renewable energy source during transmission line faults are shown.} Our examination demonstrates the effectiveness of the designed IS-ANTW compensator in reducing current deviations from constraints, thereby enhancing the resiliency of grid-connected renewable energy resources.

The remainder of the paper is structured as follows: In Section 2, \textcolor{black}{the problem is formulated from the control point of view}. Section 3 presents the LFT formulation of the output anti-windup problem, enabling its solution as a frequency-domain $H_{\infty}$ optimization. Subsequent subsections introduce the LMI formulation of the same problem to synthesize the constant gain and dynamic compensators, as detailed in Theorems 1 and 2, respectively. Section 4 outlines the scheme for joint compensation of the state and input using the LFT formulation, employing two different methods leading to full-matrix compensation and the fixed-structure approach. Section 5 presents simulation results for both a mathematical dynamic system example and a grid-connected inverter-based resource (IBR) experiencing overcurrents. Finally, Section 6 provides the paper's conclusion.

\section{Problem Statement}

\begin{figure}[t]
    \centering

    \psfrag{K}[cc][cc][1][0]{$K$}
    \psfrag{-}[cc][cc][0.8][0]{$-$}
    \psfrag{+}[cc][cc][0.8][0]{$+$}
    \psfrag{Gmy}[cc][cc][1][0]{$G_{mx}$}
    \psfrag{rr}[cc][cc][1][0]{$r$}
    \psfrag{g}[cc][cc][1][0]{$G$}
    \psfrag{yy}[cc][cc][1][0]{$y$}
    \psfrag{InANTW}[cc][cc][0.7][0]{$\textbf{Input Anti-windup}$}
    \psfrag{S}[cc][cc][1][0]{$S$}
    \psfrag{P}[cc][cc][0.7][0]{$\textbf{Plant}$}
    \psfrag{C}[cc][cc][0.7][0]{$\textbf{Controller}$}
    \psfrag{uu}[cc][cc][1][0]{$u$}
    \psfrag{un}[cc][cc][1][0]{$u_{c}$}
    \psfrag{R}[cc][cc][1][0]{$R$}
    \psfrag{uh}[cc][cc][1][0]{$\hat{u}$}
    \psfrag{eu}[cc][cc][0.8][0]{}
    \includegraphics[width=0.49\textwidth]{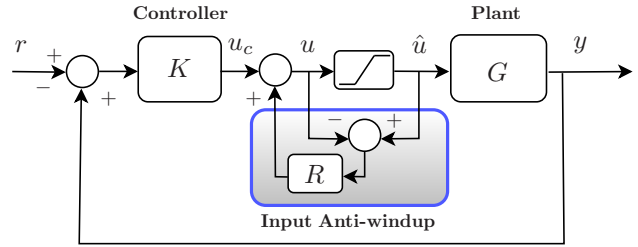}
    \caption{Structure of anti-windup compensation with hard constraints on control input.}
    \label{fig1-1}
\end{figure}

\begin{figure}[t]
    \centering

    \psfrag{K}[cc][cc][1][0]{$K$}
    \psfrag{-}[cc][cc][0.8][0]{$-$}
    \psfrag{+}[cc][cc][0.8][0]{$+$}
    \psfrag{Gmy}[cc][cc][1][0]{$G_{mx}$}
    \psfrag{w}[cc][cc][1][0]{$w$}
    \psfrag{y}[cc][cc][1][0]{$x$}
    \psfrag{y'}[cc][cc][1][0]{$\hat{x}$}
    \psfrag{umy}[cc][cc][1][0]{$u_{mx}$}
    \psfrag{Gmy}[cc][cc][1][0]{$G_{mx}$}
    \psfrag{rr}[cc][cc][1][0]{$r$}
    \psfrag{g}[cc][cc][1][0]{$G$}
    \psfrag{InANTW}[cc][cc][0.75][0]{$\textbf{State Anti-windup}$}
    \psfrag{u}[cc][cc][1][0]{$u$}
    \psfrag{r}[cc][cc][1][0]{$r$}
    \psfrag{C}[cc][cc][0.7][0]{$\textbf{Controller}$}
     \psfrag{P}[cc][cc][0.7][0]{$\textbf{Plant}$}
    \psfrag{uu}[cc][cc][1][0]{$u_{c}$}
    \psfrag{un}[cc][cc][1][0]{$u_c$}
    \psfrag{R}[cc][cc][1][0]{$R$}
    \psfrag{uh}[cc][cc][1][0]{$\hat{u}$}
    \psfrag{eu}[cc][cc][0.8][0]{}
    \includegraphics[width=0.49\textwidth]{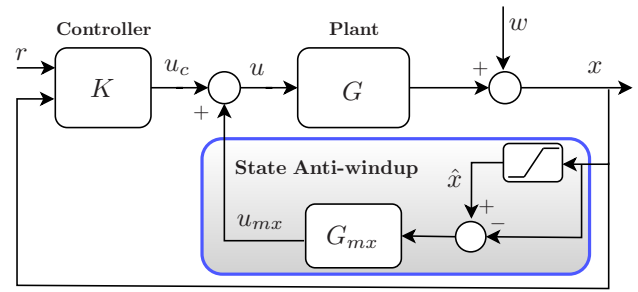}
    \caption{Proposed structure of state anti-windup compensation with soft-hard constraints on states.}
    \label{fig1-2}
\end{figure}

Fig. \ref{fig1-1} illustrates a simple classical input anti-windup structure, imposing hard constrainst on the actuator to restrict the control input $u$ during operation. It integrates an anti-windup solution, in order to condition the control input. More sophisticated methods are also present in the literature which condition both the control input and  the reference value. The shown approach solely addresses violations of the control signal $u$, from its admissible bounds, leaving no provision for imposing hard constraints on the state variable $x$. In response to this limitation, we propose the new structure depicted in Fig. \ref{fig1-2}, inspired by hard constraints, achieved through the incorporation of a saturation block within the compensator. As observed from Fig. \ref{fig1-2}, notably, the saturation block forms a key component of this limiter. The state value is passed through a saturation block and the difference between the saturated output/state and its actual value, usually referred to as ‘\textit{saturation error}’, is fed to the SANTW compensator $G_{mx}$. The outcome is then utilized for conditioning of the control input that is applied to the plant. The state anti-windup design problem aims to minimize the level of saturation  $\hat{x}-x$. However, in general case, it is also required that the level of the intervention by the state anti-windup compensator is limited. This means to regulate also the level of the signal $u_{mx}$ represented in Fig. \ref{fig1-2}. In summary, the formulation of the SANTW design problem is as follows:

\textit{\textbf{SANTW problem:} Given the plant \( G \) and the nominal controller \( K \), the objective is to find the SANTW compensator \( G_{mx} \) such that the impact of the saturated state \( \hat{x} \) and the nominal control \( u_c \) on the state constraint violation \( \hat{x} - x \) is minimized, while ensuring  a minimal intervention \( u_{mx} \) by the anti-windup compensator}. 

\textcolor{black}{Similar to classic input anti-windups, the controller K is synthesized for nominal (i.e., without saturation) conditions and constitutes input data for the SANTW problem.} Furthermore, if the state trajectory remains within the admissible set, no compensation action is undertaken by the SANTW block. \textcolor{black}{SANTW continuously monitors state constraint violations and responds accordingly. It acts as a real-time \textit{protection} based on state measurements rather than on the model, and in this important aspect, it differs from the typical soft constraints included in the compensator synthesis process. However, it is also not a hard constraint-based method like input anti-windup, where the limited input is applied directly to the plant. For these reasons, it is referred to as a \textit{soft-hard} constraint approach in the sequel.}
In this paper, the compensation is achieved through additive conditioning of the control input $u$. It is worth noting that conditioning of the reference may also be performed. However, this paper primarily focuses on input conditioning, while the discussion on reference conditioning will be addressed in future work.



In what follows we formulate the problem of state anti-windup design using two frameworks. Firstly, the problem is formulated in frequency-domain as a $H_{\infty}$ norm minimization problem which can be solved using standard tools. Secondly, the same problem is formulated based on the state-space representation and convex optimizations via LMIs. While the proposed problem is being tackled using an $H_{\infty}$ norm minimization approach it is important to note that the problem transcends this particular methodology. Alternative formulations, beyond the scope of $H_{\infty}$ , may also be explored to address the complexity and nuances of the problem more comprehensively. For instance, other methods like mixed $H_2$/$H_{\infty}$, $\mu$-synthesis, or even nonlinear optimization techniques may offer other potentially effective solutions.

\section{State/Output Anti-windup: frequency-domain formulation}

\textcolor{black}{The configuration shown in Fig. \ref{fig1-2} comprises two loops: the outer nominal control loop and the inner anti-saturation loop. As for the classic input antiwindpus, we assume that the outer loop for reference tracking is tuned in a first stage for asymptotic stability. The obtained nominal regulator K is thus input data for the design of the inner loop.} This inner loop is designed to mitigate the effects of the state saturation through control-input conditioning.

In this section we formulate the state anti-windup introduced in the previous section into the standard LFT representation which then make it possible to solve the norm minimization using standard frequency-domain tools. 

\begin{figure}[t!]
    \centering
    \psfrag{K}[cc][cc][1][0]{$K$}
    \psfrag{-}[cc][cc][0.8][0]{$-$}
    \psfrag{+}[cc][cc][0.8][0]{$+$}
    \psfrag{Gmy}[cc][cc][1][0]{$G_{my}$}
    \psfrag{w}[cc][cc][1][0]{$\hat{w}$}
    \psfrag{g}[cc][cc][1.2][0]{$G$}
    \psfrag{y'}[cc][cc][1][0]{$\hat{y}$}
    \psfrag{y}[cc][cc][1][0]{$y$}
    \psfrag{r}[cc][cc][1][0]{$r$}
    \psfrag{uc}[cc][cc][1][0]{$u_{c}$}
    \psfrag{u}[cc][cc][1][0]{$u$}
    \psfrag{w}[cc][cc][1][0]{$w$}
    \psfrag{e}[cc][tt][1][0]{$u_c$}
    \psfrag{umy}[cc][cc][1][0]{$u_{my}$}
    \includegraphics[width=0.35\textwidth]{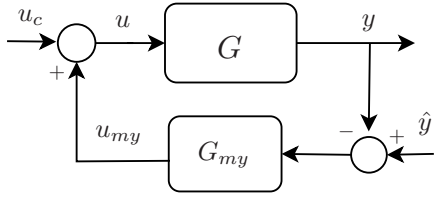}
    \caption{Output-limiter modification loop using OANTW}
    \label{fig2}
\end{figure}

For the sake of simplicity and due to space limitations, we assume zero external disturbance on the state/output ($w=0$ in Fig.  \ref{fig1-2}) for the remainder of this section. However, it is important to note that in all proposed methods, it will be straightforward to incorporate the effects of such disturbances into the obtained results with minor modifications. Furthermore, in the context of frequency-domain analysis, we utilize the output notation $y$ instead of $x$. It is essential to note that the problem remains fundamentally similar for both state and output, provided that appropriate transfer functions, ${G_{mx}}(s)$ or ${G_{my}}(s)$, are defined in each case. Thus, for now we deal with $y$, $\hat{y}$, $u_{my}$, $G_{my}$ instead of $x$, $\hat{x}$,$u_{mx}$, $G_{mx}$, respectively.
Fig. \ref{fig2} shows the output anti-windup (OANTW) loop as a part of the closed-loop structure in Fig. \ref{fig1-2}. Let us define the anti-windup sensitivity function as
\begin{equation}
    S=\left({I+G{{G}_{my}}}\right)^{-1}.
    \label{eq1}
\end{equation}
As a result,
\begin{equation}
    \hat{y}-y=S\hat{y}-SGu,
    \label{eq2}
\end{equation}
To reduce the saturation error $\hat{y}-y$ effectively, one might consider setting $G_{my}$ to be a high gain. Yet, excessive corrective action could compromise the performance of the reference tracking loop or potentially induce input/actuator saturation. Hence, there is a preference to maintain the conditioning action at a minimal level, \textcolor{black}{as already stated in the SANTW problem formulation in Section 2}. The corrective action is calculated as
\begin{equation}
    {{u}_{my}}={{G}_{my}}\left( \hat{y}-y \right)={{G}_{my}}S\hat{y}-{{G}_{my}}SGu_c,
    \label{eq3}
\end{equation}

\begin{figure}[t!]
    \centering
    \psfrag{K}[cc][cc][1][0]{$K$}
    \psfrag{-}[cc][cc][0.8][0]{$-$}
    \psfrag{+}[cc][cc][0.8][0]{$+$}
    \psfrag{Gmy}[cc][cc][1][0]{$G_{my}$}
    \psfrag{g}[cc][cc][1.2][0]{$G$}
    \psfrag{y'-y}[cc][cc][1][0]{$\hat{y}-y$}
    \psfrag{y}[cc][cc][1][0]{$y$}
    \psfrag{r}[cc][cc][1][0]{$r$}
    \psfrag{uc}[cc][cc][1][0]{$u_{c}$}
    \psfrag{u}[cc][cc][1][0]{$u$}
    \psfrag{w}[cc][cc][1][0]{$\tilde{w}$}
    \psfrag{e}[cc][cc][1][0]{$e$}
    \psfrag{z2t}[cc][cc][1][0]{${\tilde{z}}_2$}
    \psfrag{P}[cc][cc][1.2][0]{$P$}
    \psfrag{z}[cc][cc][1][0]{$z$}
    \includegraphics[width=0.35\textwidth]{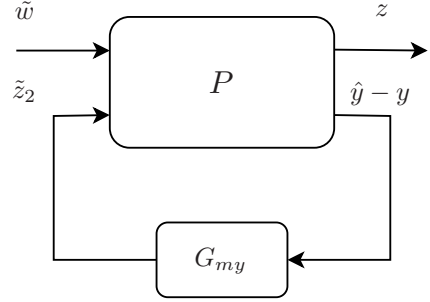}
    \caption{Anti-windup Loop.}
    \label{fig3}
\end{figure}

It is desired that the energy transfer from the disturbance inputs, which are here considered as  the saturated output $\hat{y}$, and the nominal control input $u_c$ is minimized. Therefore, \textcolor{black}{the OANTW problem is translated into}

\begin{equation}
\underset{stabilizing\,{{G}_{my}}}{\mathop{Min}}\,\frac{{{\left\| {\tilde{z}} \right\|}_{2}}}{{{\left\| {\tilde{w}} \right\|}_{2}}},\,\,\,\,\,\,\tilde{z}=\left[ \begin{matrix}
   \hat{y}-y  \\
   {{u}_{my}}  \\
\end{matrix} \right],\,\,\tilde{w}=\left[ \begin{matrix}
   {\hat{y}}  \\
   u_c  \\
\end{matrix} \right],
    \label{eq4}
\end{equation}
where $\tilde{z}$ is the vector of the outputs to be controlled and $\tilde{w}$ is the vector of disturbance inputs. 
Unlike most literature on (input) anti-windup, which defines disturbance rejection as minimizing the effect of external disturbances on the output in the presence of input saturation, this work redefines the problem. Here, the focus is on directly reflecting the effects of saturation and the nominal controller on the violation levels of states from their admissible bounds and the degree of anti-windup action.
The minimization problem (\ref{eq4}) can be reformulated as a mixed-sensitivity problem, expressed as follows:
\begin{equation}
\underset{stabilizing\,{{G}_{my}}}{\mathop{Min}}\,{{\left\| \begin{matrix}
   {{W}_{1}}S & -{{W}_{1}}SG  \\
   {{W}_{2}}{{G}_{my}}S & -{{W}_{2}}{{G}_{my}}SG  \\
\end{matrix} \right\|}_{\infty }},\,\,\,\,\,\
    \label{eq5}
\end{equation}
Here, $W_1(s)$ and $W_2(s)$ represent suitable weighting functions that capture the frequency characteristics of the desired outputs. Additionally, the problem can be transformed into a standard LFT representation using the structure depicted in Fig. \ref{fig3} where $z_1$ and $z_2$ are defined as the weighted outputs and the open-loop plant $P$ is represented as
\begin{equation}
\left[ \begin{matrix}
   {{z}_{1}}  \\
   {{z}_{2}}  \\
   \hat{y}-y  \\
\end{matrix} \right]=\left[ \begin{matrix}
   {{W}_{1}} & -{{W}_{1}}G & -{{W}_{1}}G  \\
   0 & 0 & {{W}_{2}}  \\
   I & -G & -G  \\
\end{matrix} \right]\left[ \begin{matrix}
   {\hat{y}}  \\
   {{u}_{c}}  \\
   {{{\tilde{z}}}_{2}}  \\
\end{matrix} \right]
    \label{eq6}
\end{equation}
Therefore, the minimization becomes
\begin{equation}
Mi{{n}_{{{G}_{m}}}}{{\mathcal{F}}_{l}}{{\left( P,{{G}_{m}} \right)}_{\infty }}
    \label{eq7}
\end{equation}

The minimization problem can be easily tackled using standard optimization tools. Moreover, the proposed plan can seamlessly transition to utilizing the state vector instead of the output variable, requiring no adjustments. This facilitates the straightforward adoption of the concept of state anti-windup (SANTW).

While frequency-domain $H_{\infty}$ synthesis offers robustness and performance optimization advantages, it also presents challenges such as computational complexity, conservatism, and tuning difficulties. Additionally, the resulting compensator tends to be high-order posing a challenge for designing fixed-order controllers without employing model-order reduction techniques. In the subsequent section, the problem is reformulated in terms of Linear Matrix Inequalities (LMIs). Specifically, in Section 4.2, the same dynamic SANTW compensator designed here using a frequency-domain approach will be redesigned using LMI techniques.

\section{State Anti-windup: LMI formulation}

In this section, we revisit the problem introduced in Fig. \ref{fig1-2}. Firstly, we investigate the feasibility of a constant SANTW compensator, aiming for simplicity and ease of implementation. Such a compensator, characterized by a constant-gain matrix ($G_{mx}=K_m$), while easy to design using LMI techniques, is not typically addressed by the standard frequency-domain $H_{\infty}$ method.
In the next subsection, we aim to design the dynamic SANTW compensator using an LMI approach. This approach leverages the benefits of convex optimization, allowing for the formulation of different control objectives beyond $H_{\infty}$ norm minimization. These objectives may include criteria such as D-stability, regional input restriction, parameter uncertainties, and even time-delay control. In current paper, we present the main results on the LMI formulation of the static and dynamic SANTW design and defer exploration of additional methodologies to future research endeavors.    

\subsection{LMI formulation of SANTW design: Case of constant gain}
Assume the SANTW compensator to be constant gain matrix $K_m$. The state-space equations of the system including SANTW becomes:
\begin{align}
  & \dot{x}=Ax+B{u_c}+B{{K}_{m}}\left( \hat{x}-x \right), \nonumber \\ 
 & {{z}_{1}}=-x+\hat{x}, \nonumber \\ 
 & {{z}_{2}}={{K}_{m}}{{z}_{1}}=-{{K}_{m}}x+{{K}_{m}}\hat{x},
 \label{eq8}
\end{align}
The optimization problem (\ref{eq4}) is written in this case
\begin{equation}
\underset{stabilizing\,{{K}_{m}}}{\mathop{Min}}\,\frac{{{\left\| z \right\|}_{2}}}{{{\left\| w \right\|}_{2}}},\,\,\,\,\,\,z=\left[ \begin{matrix}
   \hat{x}-x  \\
   {{u}_{mx}}  \\
\end{matrix} \right],\,\,w=\left[ \begin{matrix}
   {\hat{x}}  \\
   u_c  \\
\end{matrix} \right],
    \label{eq9}
\end{equation}
where $u_{mx}={K_m}(\hat{x}-x)$  is the compensation action by the constant-gain SANTW.   
It is possible to add some flexibility to the problem via attributing different weights on inputs/outputs. Thus, the more general case of the optimization problem will be finding $K_m$ such that
\begin{equation}
\alpha \left\| {{u}_{mx}} \right\|_{2}^{2}+\beta \left\| \hat{x}-x \right\|_{2}^{2}<{{\gamma }_{1}}\left\| {\hat{x}} \right\|_{2}^{2}+{{\gamma }_{2}}\left\| {{u}_{c}} \right\|_{2}^{2},
    \label{eq10}
\end{equation}
The subsequent theorem presents the Linear Matrix Inequality (LMI) condition to ensure the internal stability of the SANTW loop and to meet the disturbance attenuation criterion.\\

\textbf{Theorem 1:} Suppose there exist some positive  definite matrix $Q$, the matrix $Y$ and scalars $\alpha, \beta, \gamma_1, \gamma_2$ such that for some appropriately selected positive definite matrix $Q_c$, the following LMI conditions hold
\begin{align}
\label{eq12}
  & J=\left[ \begin{matrix}
   {{J}_{11}} & B & BY & -\beta Q & -\alpha {{Y}^{T}} \nonumber \\
   * & -{{\gamma }_{2}}I & 0 & 0 & 0 \nonumber  \\
   * & * & {{J}_{33}} & \beta Q & \alpha {{Y}^{T}}  \nonumber \\
   * & * & * & -\beta I & 0  \nonumber \\
   * & * & * & * & -\alpha I \nonumber  \\
\end{matrix} \right]<0, \nonumber \\ 
\end{align}
where
\begin{align}
& {{J}_{11}}=AQ+Q{{A}^{T}}-BY-{{Y}^{T}}{{B}^{T}},\,\, \nonumber \\ 
& {{J}_{33}}=-{{\gamma }_{1}}\left( {{Q}^{T}}{{Q}_{c}}+Q_{c}^{T}Q-Q_{c}^{T}{{Q}_{c}} \right)\,\,,\nonumber
\end{align}

Then: the static SANTW guarantees the  stability of the anti-windup loop 
, and  the disturbance attenuation condition (\ref{eq10}) is satistfied. Finally, the compensator gain $K_m$ is calculated as $K_m=YQ^{-1}$.

\textit{Proof}: The proof is provided in Appendix A. \\
\qed \qed \qed \\
Finding the matrix $Q_c$ can be challenging here. For simplifying this process, we propose Algorithm (\ref{A1}) which is an iterative way to find the needed matrix $Q_c$.
\begin{algorithm}[t!]
	\caption{Iterative solution of (\ref{eq12})}
	\label{A1}
	\begin{algorithmic}[1]
		\State For a set of parameters $\alpha, \beta, {\gamma}_2, {\bar{u}}_i, i=1,...,m,$ find a $Q_c>0$ such that the LMI (\ref{eq12}) is feasible.
		\State Store the solution of $Q$, ${\gamma}_1$ obtained from the LMI.
		\State Decrease ${\gamma}_1$ to the possible extent with the same $Q_c$.
		\State If $\|Q-Q_c\|>\varepsilon$, Put $Q_c=Q$ and go back to step 2.
		\State Put $K_m=YQ^{-1}$.
	\end{algorithmic}
\end{algorithm}

\textit{Remark 1:} Putting different weights on $z_1, z_2$ or considering ${\gamma}_1$, ${\gamma}_2$ instead of only ${\gamma}$ can be useful in synthesizing a satisfactory compensator. \\ 
\textit{Remark 2:} The step 1 in the algorithm can be still challenging. To find an initial value for $Q_c$ , one way can be generating positive definite matrices of the appropriate dimension in a random loop. \\ 
\textit{Remark 3:} Rather than the static weights, one can design dynamic weights $W_1(s)$, $W_2(s)$ to have different performance characteristics in different frequencies. \\

The design process employing Theorem 1 and Algorithm 1 proves to be highly effective and computationally efficient. However, the compensator designed using this method is a static gain, which, while easy to implement, may not yield the same performance efficiency as designing $n$ dynamic compensators one for each of the saturated states. 

\subsection{LMI formulation of SANTW design: Case of dynamic compensator}
In this section, we explore the design of a dynamic SANTW compensator using an LMI formulation. Previously, in Section 3, we designed a dynamic compensator based on frequency-domain $H_{\infty}$ optimization, which relies on Riccati equations. However, the motivation for employing the LMI formulation lies in its reliability and computational efficiency in solving convex optimization problems. Additionally, the LMI formulation opens pathways for incorporating other time-domain control objectives in future endeavors. Examples of such objectives could include addressing problems like D-stability and $H_2$ control (\cite{duan2013lmis, abolmasoumi2011robust}).

Designing a dynamic compensator based on LMIs is significant for another reason.  In some anti-windup studies, both the reference and control input are modified (\cite{weston2000linear},\cite{tarbouriech2011stability} and \cite{zaccarian2011modern}). In such compensation, when both conditioning gains are constant, it can be demonstrated equivalent to designing a single \textit{dynamic} conditioning applied only on the control input.

Let us reconsider the proposed structure in Fig. \ref{fig1-2} and let a state-space realization to the SANTW compensator $G_{mx}(s)$ be ($A_q,B_q,K_{m2},K_{m1}$).
A generalized form of equation (\ref{eq8}) in presensce of the dynamic SANTW compensator is

\begin{align}
\label{eq13}
  & \dot{x}=Ax+B{{u}_{c}}+B{{K}_{m1}}\left( \hat{x}-x \right)+B{{K}_{m2}}q, \nonumber \\ 
 & \dot{q}={{A}_{q}}q+{{B}_{q}}\left( \hat{x}-x \right), \nonumber \\ 
 & {{z}_{1}}=\hat{x}-x,\nonumber \\ 
 & {{z}_{2}}={{K}_{m1}}\left( \hat{x}-x \right)+{{{{K}}}_{m2}}q, 
\end{align}
in which $q$ is the state of SANTW compensator. The anti-windup action $z_2$ is influenced both by the state of the compensator and directly by the state saturation error $\hat{x}-x$. There are four parameters to be designed: $A_q$, $B_q$, $K_{m1}$, and $K_{m2}$. However, it is often observed that the same level of performance can be achieved with fewer parameters. For instance, setting $A_q=0$ and $B_q=I$ results in a MIMO-PI compensator. For simplicity in obtaining the LMIs, we assume $B_q=I$ in the remainder of this section. The following theorem summarizes the results on designing the dynamic SANTW using the LMIs.

\textbf{Theorem 2:} 
Suppose there exist some positive definite matrix $Q_1$ , matrices $Y_1$ , $Y_2$, $Y_A$  with matching dimensions, and some positive scalars $h_1$,$h_2$, $\alpha$, $\beta$, $\gamma_1$, $\gamma_2$ such that for some appropriately selected positive definite matrix $Q_c$, the following LMI conditions hold
\begin{equation}
    \label{eq14}
    \resizebox{0.7\hsize}{!}{$
    \begin{aligned}
    \Gamma =\left[ \begin{matrix}
   {{\Gamma }_{11}}+\Gamma _{11}^{*} & {{{\tilde{B}}}_{u}} & {{\Gamma }_{13}} & {{\Gamma }_{14}}  \\
   * & -{{\gamma }_{1}}I & 0 & 0  \\
   * & * & {{\Gamma }_{33}} & {{\Gamma }_{34}}  \\
   * & * & * & {{\Gamma }_{44}}  
\end{matrix} \right]<0,
    \end{aligned} $}
\end{equation}
\begin{equation}
    \resizebox{0.9\hsize}{!}{$
    \begin{aligned}
    Q  &=\left[ \begin{matrix}
   {{Q}_{1}} & {{h}_{1}}{{Q}_{1}}  \\
   {{h}_{1}}{{Q}_{1}} & {{h}_{2}}{{Q}_{1}}  \\
\end{matrix} \right],\,\,{{h}_{2}}>h_{1}^{2}>0, \nonumber \\ 
 {{\Gamma }_{11}}  &= {\left[ \begin{matrix}
   A{{Q}_{1}}-B{{Y}_{2}} & {{h}_{1}}A{{Q}_{1}}-B\left( {{h}_{1}}{{Y}_{1}}+{{h}_{2}}{{Y}_{2}} \right)  \\
   -{{Q}_{1}}+{{h}_{1}}{{Y}_{A}} & -{{h}_{1}}{{Q}_{1}}+{{h}_{2}}{{Y}_{A}}  \\
\end{matrix} \right],} \nonumber  
   \end{aligned}
    $}
\end{equation}
\begin{equation}
    \resizebox{0.5\hsize}{!}{$
    \begin{aligned}
    {{{\tilde{B}}}_{u}} &=\left[ \begin{matrix}
   B  \\
   0  \\
\end{matrix} \right],\,\,
{{\Gamma }_{13}}=\left[ \begin{matrix}
   B{{Y}_{1}} \nonumber \\
   {{Q}_{1}}  \\
\end{matrix} \right],\,\nonumber  
    \end{aligned}
    $}
\end{equation}
\begin{equation}
    \resizebox{0.7\hsize}{!}{$
    \begin{aligned}
    {{\Gamma }_{14}} &=\left[ \begin{matrix}
   -\beta {{Q}_{1}} & \alpha \left( -Y_{1}^{T}+{{h}_{1}}Y_{2}^{T} \right)  \\
   -\beta {{h}_{1}}{{Q}_{1}} & \alpha \left( -{{h}_{1}}Y_{1}^{T}+{{h}_{2}}Y_{2}^{T} \right)  \\
\end{matrix} \right], \nonumber
   \end{aligned}
    $}
\end{equation}
\begin{equation}
    \resizebox{0.7\hsize}{!}{$
    \begin{aligned}
    {{\Gamma }_{33}} &=-{{\gamma }_{2}}\left( Q_{1}^{T}{{Q}_{c}}+Q_{c}^{T}{{Q}_{1}}-Q_{c}^{T}{{Q}_{c}} \right),\nonumber \\ 
  {{\Gamma }_{34}} &=\left[ \begin{matrix}
   \beta {{Q}_{1}} & \alpha Y_{1}^{T}  \\
\end{matrix} \right],\,\,{{\Gamma }_{44}}=\left[ \begin{matrix}
   -\beta I & 0 \nonumber \\
   0 & -\alpha I  \\
\end{matrix} \right],\nonumber
    \end{aligned}
    $}
\end{equation}

Then the anti-windup loop is internally stable and the disturbance attenuation condition (\ref{eq10}) is satisfied. 
The controller gains are then calculated as $K_{m1}={Y_1}{Q_1}^{-1}$, $K_{m2}={Y_2}{Q_1}^{-1}$, and $A_{q}={Y_A}{Q_1}^{-1}$.
\textit{Proof}: The proof is provided in Appendix B. \\
\qed \qed \qed \\

\textit{Remark 4}: It is suggested to put different weights on $z_1$, $z_2$ to find a balance between the SANTW action and the needed reduction on the state saturation. In case of the need for further improvements, the dynamic gains may also be utilized to provide different weights on different frequencies. \\
\textit{Remark 5}: It is necessary to use Algorithm 1 to find the best values for the disturbance attenuation level $\gamma$ (or ${\gamma}_1,{\gamma}_2$ in case of various attenuation levels).

\section{Joint State-Input anti-windup compensation}
Control saturation, as discussed earlier, is inherent in many practical dynamic systems. This saturation is often considered a strict constraint on the control input. Ignoring these constraints in SANTW design can lead to performance degradation or system oscillations. In the simulation section, it is shown that applying only SANTW compensation results in a spike-like behavior in the control input (Figures \ref{fig8} and \ref{fig12}, illustrating the control input required for SANTW in the cases of frequency-domain and LMI-based $H_{\infty}$, respectively). Despite efforts such as using the Lyapunov region method, applying penalty weights on anti-windup action, and attempting to restrict the SANTW action, the spike-like behavior persists.

In order to address such challenge, we propose a novel scheme for addressing the input saturation and state saturation at the same time. Figure \ref{fig4} illustrates the proposed joint input and state anti-windup (IS-ANTW) compensation plan. It shows two interconnected loops: one for compensating output (or state) saturation and the other for conditioning the error signal to protect the control input from saturation. The objective of this structure is to minimize the impact of disturbances, including the error signal $e$, saturated input $\hat{u}$, and saturated output $\hat{y}$, on the controlled outputs, including saturation errors on the output and control input, i.e., $y-\hat{y}$ and $u-\hat{u}$, respectively. As for SANTW, this is formulated as the following distrurbance rejection problem:

\begin{figure}[t!]
    \centering
    \psfrag{aK}[cc][cc][1][0]{$K$}
    \psfrag{a-}[cc][ll][0.8][0]{$-$}
    \psfrag{a+}[cc][cc][0.8][0]{$+$}
    \psfrag{aGmy}[cc][cc][1][0]{$G_{my}$}
    \psfrag{aGmu}[cc][cc][1][0]{$G_{mu}$}
    \psfrag{aw}[cc][cc][0.8][0]{$\hat{w}$}
    \psfrag{ag}[cc][cc][1.2][0]{$G$}
    \psfrag{ayh}[cc][tt][1][0]{$\hat{y}$}
    \psfrag{auhat}[cc][cc][0.8][0]{$\hat{u}$}
    \psfrag{y}[cc][cc][1][0]{$y$}
    \psfrag{ae}[cc][cc][0.8][0]{$e$}
    \psfrag{aeu}[cc][cc][1][0]{$e_u$}
    \psfrag{aumu}[cc][cc][0.8][0]{$u_{mu}$}
    \psfrag{aumy}[cc][cc][0.8][0]{$u_{my}$}
    \psfrag{au}[cc][cc][0.8][0]{$u$}
    \psfrag{pl}[cc][cc][0.6][0]{$\textbf{Plant}$}
    \psfrag{nc}[cc][cc][0.6][0]{$\textbf{Nominal Controller}$}
    \psfrag{Ov}[cc][cc][0.8][0]{$\textbf{IS-ANTW Controller}$}
    \psfrag{w}[cc][cc][1][0]{$w$}
    \psfrag{e}[cc][tt][1][0]{$u_c$}
    \psfrag{In}[cc][tt][0.7][0]{$\textbf{Input Anti-windup}$}
    \psfrag{Ou}[cc][tt][0.7][0]{\parbox{3cm}{\centering $\textbf{Output/state}$ \\ $\textbf{Anti-windup}$}}
    \includegraphics[width=0.5\textwidth]{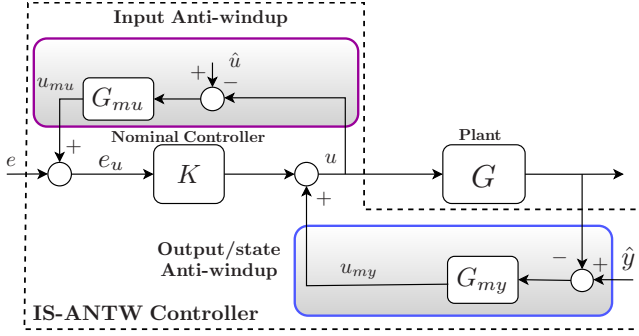}
    \caption{Joint Input/output Anti-windup Plan}
    \label{fig4}
\end{figure}

\begin{equation}
   \underset{stabilizing\,{{G}_{mu}},{{G}_{my}}}{\mathop{Min}}\,\frac{{{\left\| {\tilde{z}} \right\|}_{2}}}{{{\left\| {\tilde{w}} \right\|}_{2}}},
\label{eq15}
\end{equation}
where
\begin{equation}
\tilde{z}=\left[ \begin{matrix}
   \hat{y}-y  \\
   \hat{u}-u  \\
\end{matrix} \right],\,\,\tilde{w}=\left[ \begin{matrix}
   {\hat{y}}  \\
      e  \\
   {\hat{u}} \nonumber
\end{matrix} \right],  
\end{equation}

\textcolor{black}{Structure in Fig. \ref{fig4} and formulation (\ref{eq15}) are now analysed.} We first need to establish the relationship between the current saturation errors $y-\hat{y}$, $u-\hat{u}$, and the disturbance signals. The output saturation error can be written as 
\begin{equation}
    \hat{y}-y={{S}_{y}}\hat{y}-{{S}_{y}}GKe-{{S}_{y}}GK{{G}_{mu}}\left( \hat{u}-u \right),
    \label{eq16}
\end{equation}
For ${{S}_{y}}={{\left( I+G{{G}_{my}} \right)}^{-1}}$. Similarly, 
\begin{align}
  & \hat{u}-u={{S}_{u}}\hat{u}-{{S}_{u}}Ke-{{S}_{u}}{{u}_{my}}, \nonumber \\ 
 & {{S}_{u}}={{\left( I+K{{G}_{mu}} \right)}^{-1}},
 \label{eq17}
\end{align}
Now by substituting (\ref{eq17}) in (\ref{eq16}), the output saturation errors is obtained as
\begin{align}
    \hat{y}-y & ={{M}_{y}}{{S}_{y}}\hat{y}-{{M}_{y}}{{S}_{y}}GK\left( I-{{G}_{mu}}{{S}_{u}}K \right)e \nonumber \\
   &-{{M}_{y}}{{S}_{y}}GK{{G}_{mu}}{{S}_{u}}\hat{u},
 \label{eq18} 
\end{align}
where ${{M}_{y}}={{\left( I-{{S}_{y}}GK{{G}_{mu}}{{S}_{u}}{{G}_{my}} \right)}^{-1}}$.
Similarly, the input saturation error is calculated in terms of disturbances as
\begin{align}
    \hat{u}-u & ={{M}_{u}}{{S}_{u}}\hat{u}-{{M}_{u}}{{S}_{u}}\left( I-{{G}_{my}}{{S}_{y}}GK \right)e \nonumber \\
   &-{{M}_{u}}{{S}_{u}}{{G}_{my}}{{S}_{y}}\hat{y},
 \label{eq19}  
\end{align}
where ${{M}_{u}}={{\left( I-{{S}_{u}}{{G}_{my}}{{S}_{y}}GK{{G}_{mu}} \right)}^{-1}}$.
From (\ref{eq16}), we can deduce that achieving a small saturation error on the output requires the sensitivity function $S_y$ to be very small. This can be accomplished by selecting a large gain for the output anti-windup $G_{my}$. However, increasing $G_{my}$ also leads to an increase in $M_u$, resulting in a higher saturation error in the control input. Essentially, (\ref{eq18}) and (\ref{eq19}) illustrate a trade-off between reducing the output saturation error and the input saturation error. Therefore, with the proposed approach, balancing these two types of saturation is crucial and can be achieved through optimization. As a result, there is no need to restrict each anti-windup action individually \textcolor{black}{and optimization (\ref{eq15}) will adequately constrain the aggregate input $u$.}


\begin{figure}[t]
    \centering
    \psfrag{aK}[cc][cc][1][0]{$K$}
    \psfrag{a-}[cc][ll][0.8][0]{$-$}
    \psfrag{a+}[cc][cc][0.8][0]{$+$}
    \psfrag{Gmy}[cc][cc][0.8][0]{$\left[ \begin{matrix} {{G}_{my}} & 0  \\    0 & {{G}_{mu}} \end{matrix} \right]$}
    \psfrag{1}[1][1][1][0]{$d(t)$}
    \psfrag{P}[0][1][1.3][0]{$P$}
    \psfrag{w}[cc][ll][1][0]{$\tilde{w}$}
    \psfrag{z}[cc][cc][1][0]{$z$}
    \psfrag{z2t}[cc][cc][0.8][0]{$\left[ \begin{matrix}  {{u}_{my}}  \\   {{u}_{mu}}  \\
\end{matrix} \right]$}
    \psfrag{y'-y}[cc][rc][0.8][0]{$\left[ \begin{matrix}  {\hat{y}-y}  \\   {\hat{u}-u}  \\ \end{matrix} \right]$}
    \includegraphics[width=0.35\textwidth]{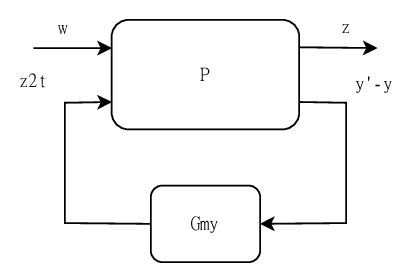}
        \caption{Joint output-input anti-windup plan with ${{\tilde{w}}={{[ \begin{matrix}
   {{{\hat{y}}}^{T}} & {{e}^{T}} & {{{\hat{u}}}^{T}}  
\end{matrix} ]}}^T, {{z}}={{[ \begin{matrix}
   z_{1}^{T} & z_{2}^{T} 
\end{matrix} ]}^{T}}.}$}

    \label{fig5}
\end{figure}

\begin{algorithm}[t!]
	\caption{Steps for designing fixed-structure input-state anti-windup compensator.}
	\label{A2}
	\begin{algorithmic}[1]
\State Design a nominal controller $K(s)$ to stabilize the closed-loop system composed of the nominal plant $G(s)$ while ensuring the desired performance.
\State Determine appropriate frequency-dependent weights $W_u(s)$ and $W_y(s)$ to address the input saturation error $\hat{u}-u$ and output/state saturation error $\hat{y}-y$, respectively.
\State Define the structure depicted in Fig. \ref{fig4} for non-smooth $H_{\infty}$ optimization.
\State Employ specialized optimization algorithms for non-smooth functions, as outlined in non-smooth $H_{\infty}$ optimization by \cite{upkarian}, to derive the compensators $G_{mu}$ and $G_{my}$.
\end{algorithmic}
\end{algorithm}

The standard representation of the joint input-output anti-windup (IO-ANTW) compensated system depicted in Fig. \ref{fig4} is shown in Fig. \ref{fig5}. By assuming the  dynamic weight functions $W_u(s)$, $W_y(s)$ to define new controlled outputs $\tilde{z}_1=W_y(s)z_1$, $\tilde{z}_2=W_u(s)z_2$  the modified plant $\tilde{P}$ is illustrated as
\begin{equation}
    \label{eq20}
    \resizebox{0.8\hsize}{!}{$
    \begin{aligned}
    \Tilde{P}\left( s \right)=\left[ \begin{matrix}
   {{W}_{y}} & -{{W}_{y}} & -{{W}_{y}}GK & 0 & -{{W}_{y}}G & -{{W}_{y}}GK  \\
   0 & 0 & -{{W}_{u}}K & {{W}_{u}} & -{{W}_{u}} & -{{W}_{u}}K  \\
   I & -I & -GK & 0 & -G & -GK  \\
   0 & 0 & -K & I & -I & -K  \\
\end{matrix} \right].
    \end{aligned}
    $}
\end{equation}

\textit{Remark 6:} It is worth noting that the diagonal form of the feedback matrix shown in Fig. \ref{fig5} might pose challenges for direct application of conventional $H_{\infty}$ optimization tools. However, this issue can be addressed by considering a full feedback matrix, allowing the non-diagonal elements to be non-zero and included in the design process. Such alternative full-matrix feedback structure is more intricate, involving the use of both the output saturation error $\hat{y}-y$ and $\hat{u}-u$ to construct the anti-windup action $u_{mu}$ and vice versa. In the sequel, such compensator will be referred to as the \textit{full-matrix input-state anti-windup} (FB-IS-ANTW).

Full-matrix compensation, however effective, causes difficulties in both design and implementation levels.
In order to retain the diagonal-matrix design, one can adopt the ‘\textit{tailored}’ specialized  $H_{\infty}$ control synthesis  proposed by \cite{upkarian}. This synthesis method does not impose any restrictions on the controller's structure or order. The algorithmic strategy leverages generalized gradients and bundling techniques tailored for meeting the $H_{\infty}$ norm criteria. In this method, descent directions are computed through quadratic programs, and steps are generated via line search. In the rest of the paper, such method is referred to as \textit{fixed-structure input-state anti-windup} (FS-IS-ANTW). \textcolor{black}{Imposing the structure of the compensator will certainly limit its performances. However, such a fixed-structure may provide the advantage of imposing a reduced order controller in case of complex plants and to avoid thus complexity in terms of implementation, tuning, and computational burden.}

The summarized steps for designing FS-IS-ANTW are presented in Algorithm 2.

\section{Simulation Results}
In what follows, the effectiveness of the proposed methods is assessed through simulations conducted on two distinct systems. The first represents a mathematical second-order system, while the second is an industrial application to the STATCOM control of a grid-connected voltage-source inverter (VSI) \cite{ngo2022emt} developed in H2020 POSYTYF project. For the latter one, Electromagnetic Transient (EMT) simulations are provided. 

\subsection{Example I: Second-order mathematical model}
Assume the following second-order system example:
\begin{equation}
   G\left( s \right)=\frac{1}{{{s}^{2}}+s+1},
   \label{eq21}
\end{equation}
A state-space realization is
\begin{align}
\label{eq22}
  & \dot{x}=Ax+Bu, \nonumber \\ 
 & y=Cx, \\ 
 & A=\left[ \begin{matrix}
   -1 & -1  \\
   1 & 0  \\
\end{matrix} \right],B=\left[ \begin{matrix}
   1  \\
   0  \\
\end{matrix} \right],C=I.  \nonumber 
\end{align}
\begin{figure}[t]
    \centering
    \includegraphics[width=0.25\textwidth]{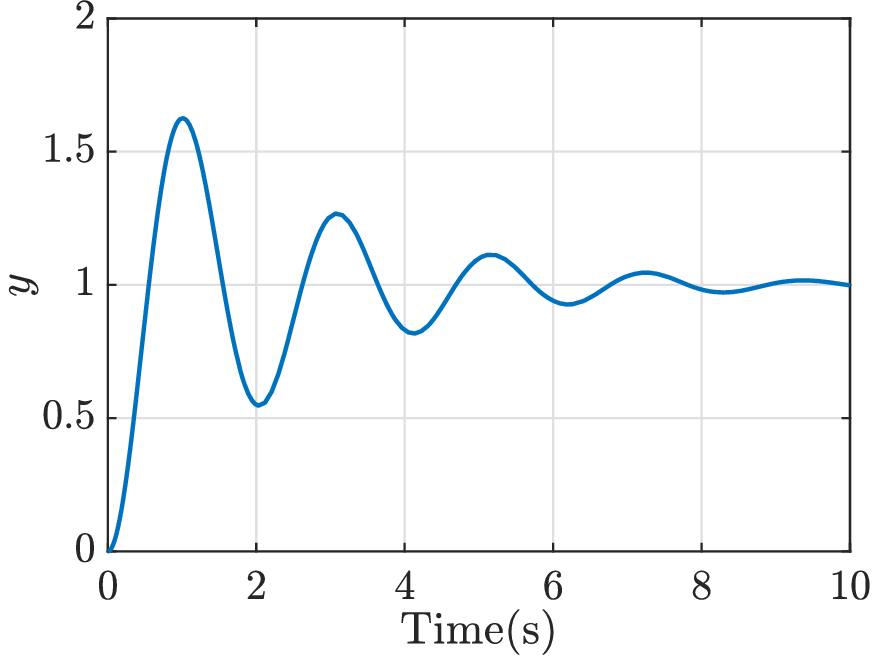}
    \caption{Nominal tracking by PID controller.}
    \label{fig6}
\end{figure} 
In the subsequent simulation, we aim to reduce oscillations in the states of the example system using various proposed SANTW methods. To effectively demonstrate the anti-windup scheme's performance, a PID controller is selected as the nominal controller and tuned to induce output tracking oscillations, as shown in Fig. \ref{fig6}. The system is supposed to follow $r=1$ as the reference. We also set one as the upper bound for the saturation which means that the constraint on state requires all states to be below one. However, this will not be completely possible without losing the nominal tracking. The designed SANTW compensators are applied to minimize the violation from such constraint on states. Four proposed design approaches are evaluated: the frequency-based dynamic $H_{\infty}$
  method, the static SANTW, the LMI-based dynamic SANTW, and the joint input-state $H_{\infty}$
(IS-ANTW).
\begin{figure}[htbp]
\centering
\subfloat[]{\includegraphics[width=0.24\textwidth]{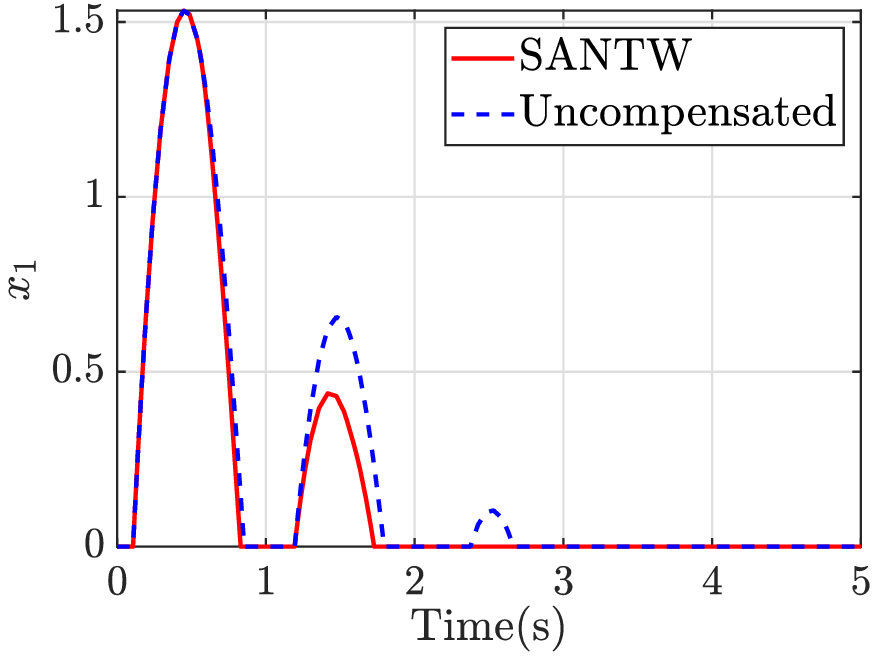}}
\subfloat[]{\includegraphics[width=0.24\textwidth]{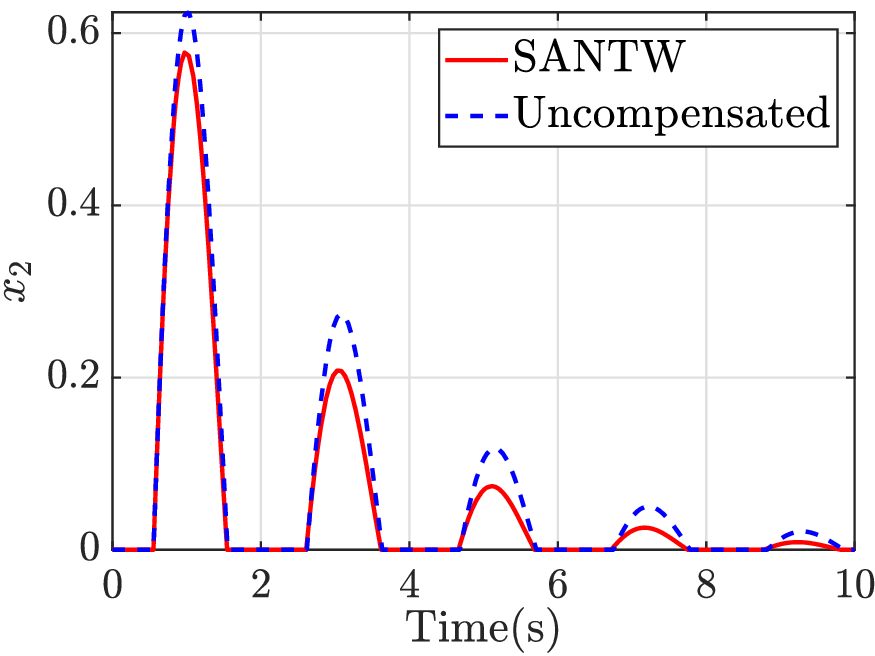}} \\
\subfloat[]{\includegraphics[width=0.24\textwidth]{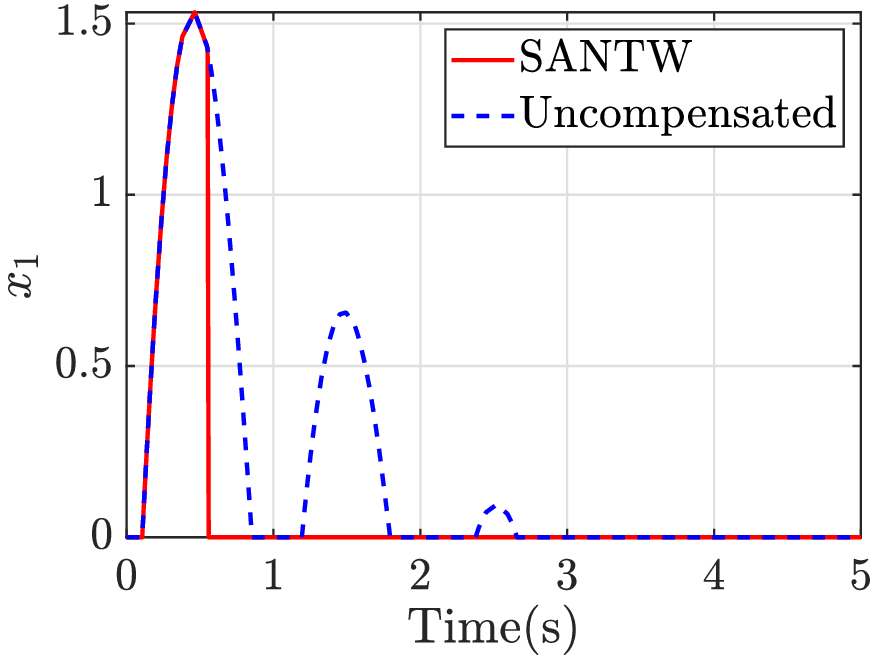}}
\subfloat[]{\includegraphics[width=0.24\textwidth]{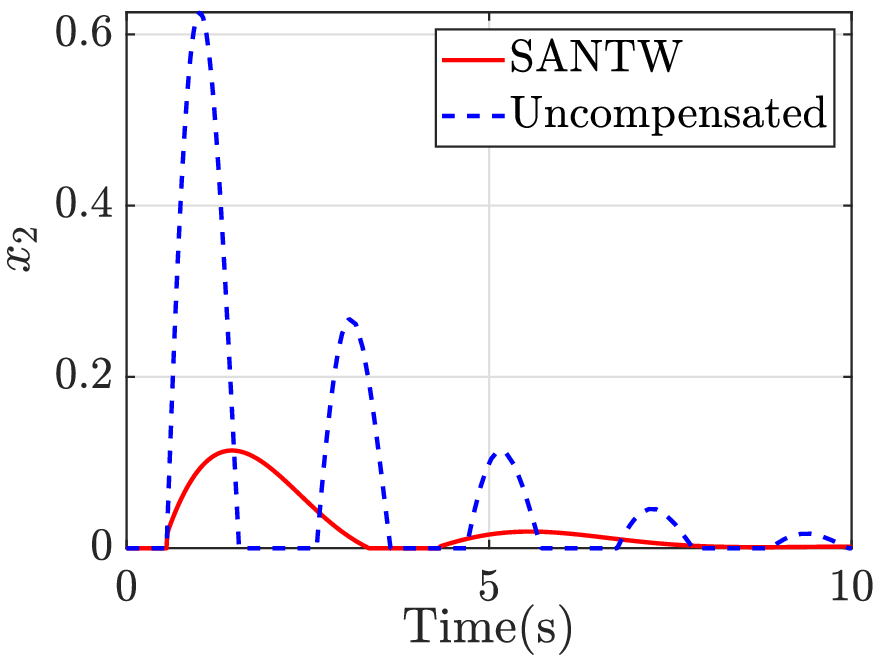}}
\caption{Performance of frequency-domain dynamic state anti-windup compensator: state saturation errors $\hat{x}-x$ for (a), (b) $W_1=10$, $W_2=0.01$, and (c),(d) for weights in (\ref{eq23}).}
\label{fig7}
\end{figure}

\textit{(a) SANTW design based on frequency-domain $H_{\infty}$ oprtimization}:  We begin by deriving the basic state anti-windup compensator through the optimization process outlined in (\ref{eq4}). Setting $C=I$ allows us to create a MIMO system where the states are utilized as output variables. The design weights, represented by $W_1$ and $W_2$, significantly influence the quality of the anti-windup compensation. Initially, we investigate the effects of using constant weights, such as $W_1=10$ and $W_2=0.01$. Next, we can choose frequency-dependent weights tailored to the oscillation frequency of the system response and the desired level of intervention have been used
\begin{equation}
\label{eq23}
    {{W}_{1}}=\frac{s+155.5}{s+15.24},\,\,\,{{W}_{2}}=0.01.
\end{equation}
The saturation errors are illustrated in Fig. \ref{fig7}. Where the impact of frequency-dependent weights on reducing state saturation errors is clearly demonstrated, i.e. violations of state constraints, termed as $\hat{x}-x$ are less when suitable weights are selected.  

\begin{figure}[t]
\centering
\subfloat[]{\includegraphics[width=0.24\textwidth]{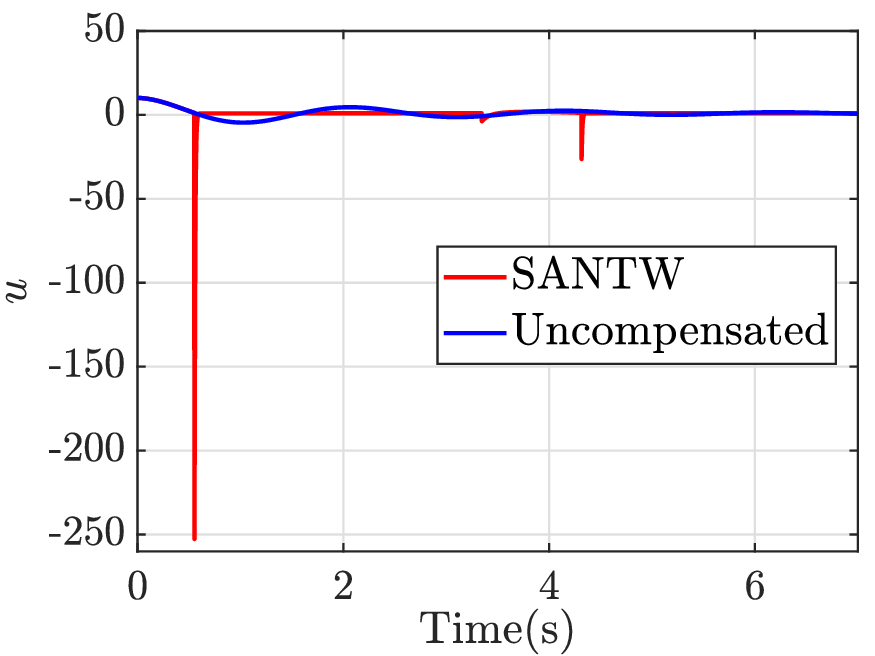}}
\subfloat[]{\includegraphics[width=0.24\textwidth]{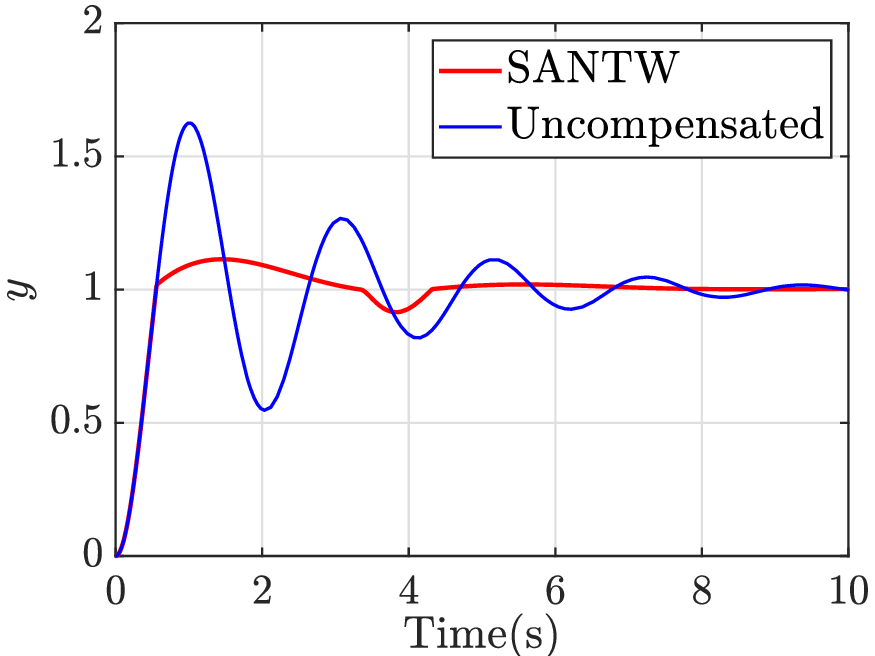}} 
\caption{The effects of the state anti-windup compensation (frequency-based $H_{\infty})$ on (a) control input and (b) closed-loop tracking behaviour.}
\label{fig8}
\end{figure}

In Fig. \ref{fig8}, a spike-like behaviour is clearly observed in the control which is the result of the SANTW action. Applying big penalties on SANTW action to avoid such behavior is not much effective. In fact, it was observed that increasing the weight $W_2$ does not notably influence the spike behavior, and further increases eventually results in a deterioration of the SANTW's performance. The presence of spike-like behavior in the compensation signal poses a challenge, particularly when applying this method to systems with actuator saturation. Additionally, the impact of integrating the SANTW compensator on the system's reference tracking behavior is illustrated in Fig. \ref{fig8}. It is promising to observe an improvement in tracking quality compared to the nominal controller. \textcolor{black}{However, the satisfactory solution can only be obtained with the joint input-state anti-windup given in part \textit{(d)} below.}

\begin{figure}[t]
\centering
\subfloat[]{\includegraphics[width=0.24\textwidth]{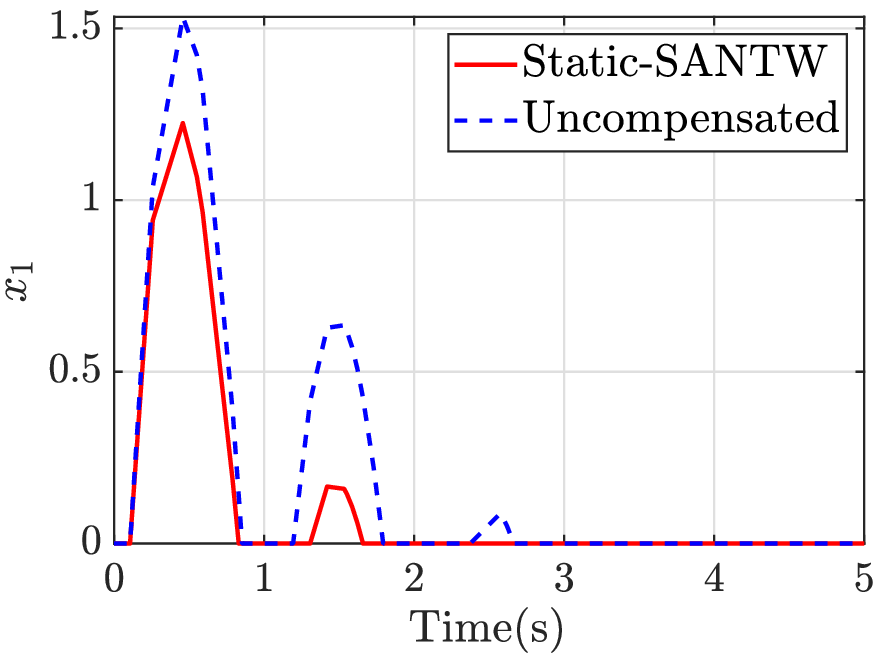}}
\subfloat[]{\includegraphics[width=0.24\textwidth]{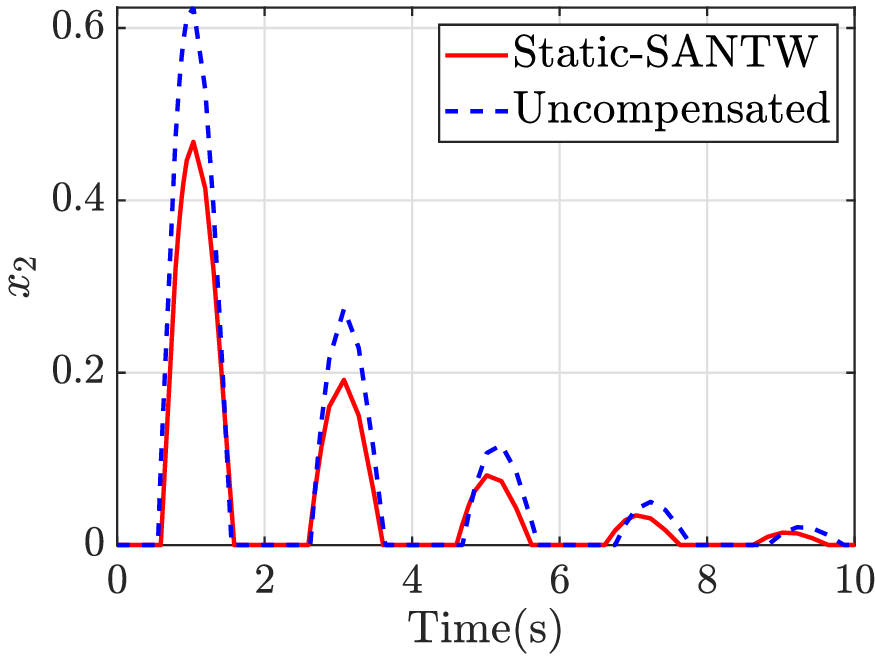}} 
\caption{Performance of the static SANTW compensator to decrease state saturation errors $\hat{x}-x$.}
\label{fig9}
\end{figure}

\begin{figure}[t]
\centering
\subfloat[]{\includegraphics[width=0.24\textwidth]{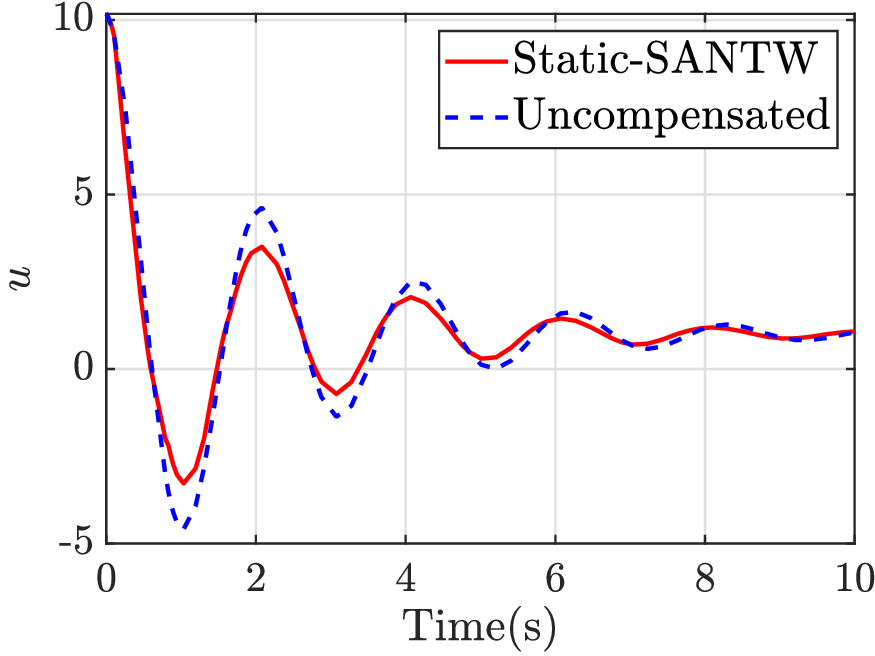}}
\subfloat[]{\includegraphics[width=0.24\textwidth]{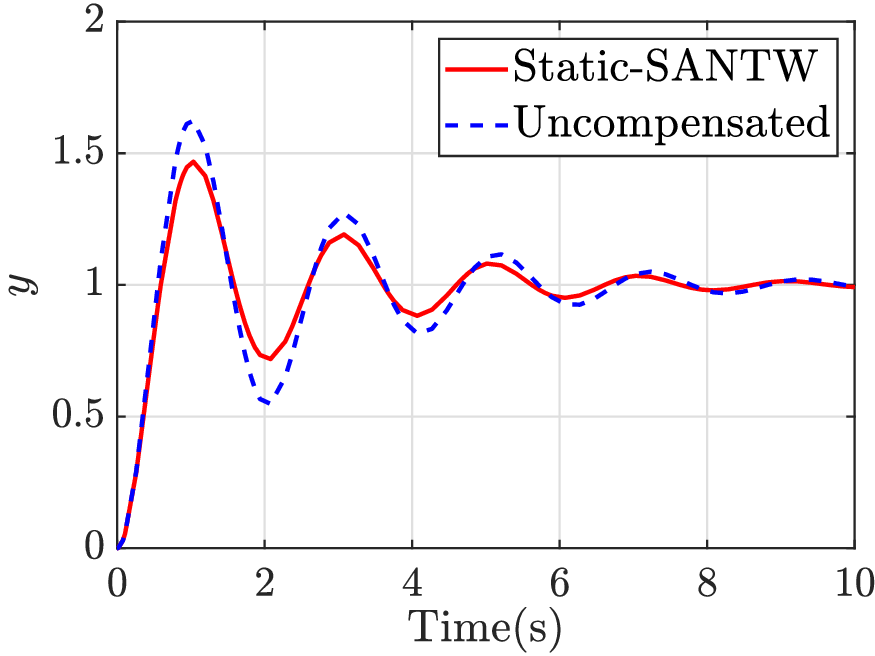}} 
\caption{The effects of the static state anti-windup compensation on (a) control input and (b) closed-loop tracking.}
\label{fig10}
\end{figure}

\textit{(b) Design of static SANTW}: 
In this part, the proposed methods in section 4.1 and Theorem 1 are examined. As mentioned before, this will brings up the ease in the computations and implementation. In an effort to solve LMI (\ref{eq12}) in Theorem 1 through Algorithm 1 , and by putting weights $\alpha=0.001$ on saturation action and the weight $\beta=3.15$ on the saturation error,  the gain is obtained as ${{K}_{m}}=\left[ 1.3057~~~0.9549 \right]$. Fig. \ref{fig9} illustrates the results on decreasing the state constraint violation caused by the oscillation which comes from improper nominal tuning of the system introduced in (\ref{eq21}), obtained by applying the static SANTW gain obtained from LMIs in Theorem 1. As observed, the results, shows improvement in state saturation. The aggregate control applied to the plant is shown in Fig. \ref{fig10}(a), indicating a decrease in control input, effectively preventing actuator saturations. However, the saturation attenuation level is not as high as with dynamic SANTW. To further reduce saturation, increasing weights on the saturation error magnitude may be considered, but this could render the design process infeasible due to challenges in finding an initial state for $Q_c$. Fig. \ref{fig10}(b) shows the comparison between the reference tracking before and after including the static anti-windup compensator. As seen, the tracking behavior is slightly improved as compared to uncompensated case.

\begin{figure}[t]
\centering
\subfloat[]{\includegraphics[width=0.24\textwidth]{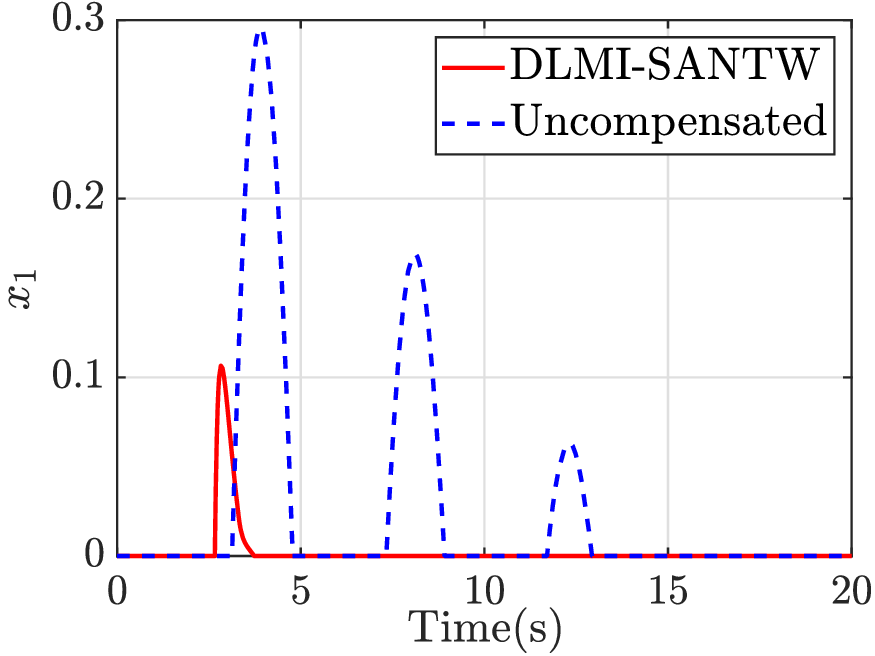}}
\subfloat[]{\includegraphics[width=0.24\textwidth]{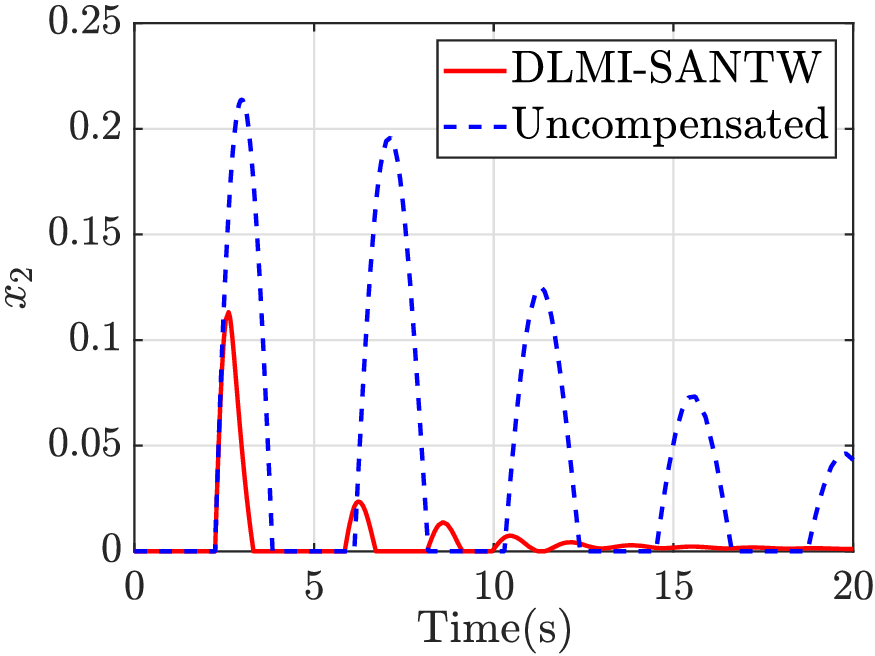}} 
\caption{Performance of the LMI-based dynamic SANTW compensator to decrease state saturation errors $\hat{x}-x$.}
\label{fig11}
\end{figure}

\begin{figure}[t]
\centering
\subfloat[]{\includegraphics[width=0.24\textwidth]{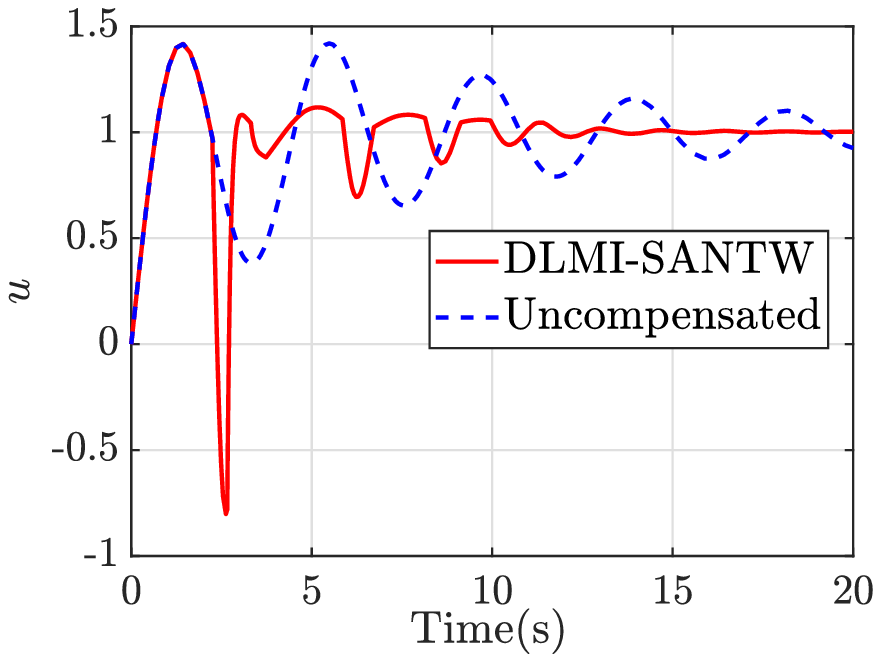}}
\subfloat[]{\includegraphics[width=0.24\textwidth]{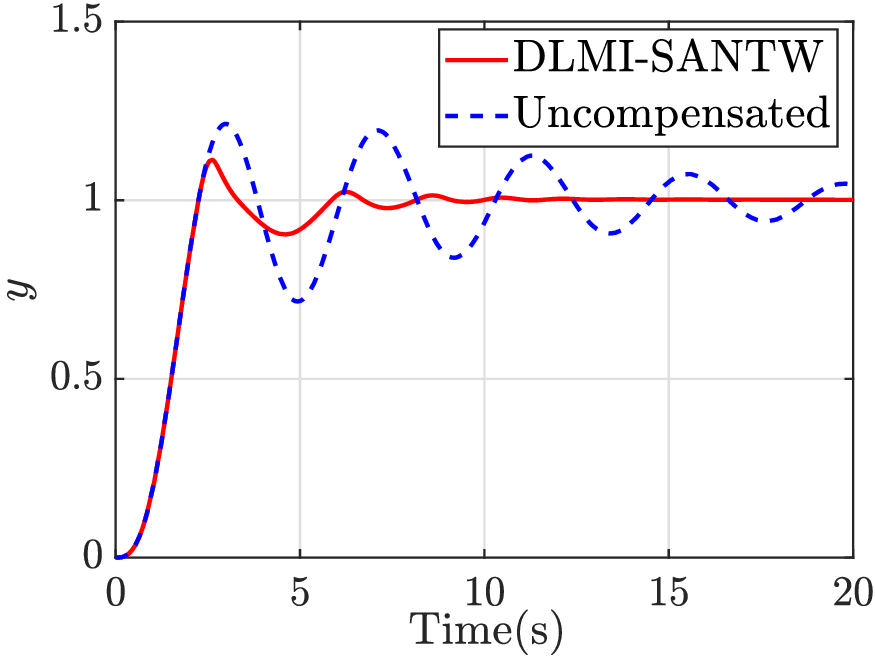}} 
\caption{The effects of the LMI-based dynamic state anti-windup compensation on (a) control input and (b) closed-loop tracking.}
\label{fig12}
\end{figure}
\textit{(c) Design of dynamic SANTW using LMIs}:
The performance of the dynamic LMI-based state anti-windup (DLMI-SANTW) compensator designed using Theorem 2 is evaluated in this part. To better demonstrate the performance of the proposed SANTW compensator, this time we employ a MIMO-PI controller (or integral tracking state feedback) as the nominal controller. The proportional and integral gains are designed using pole placement to induce some oscillation in the states. For simplicity, the assumption is made that $B_q=I$ and the tuning parameters for design are set in a manner such that an initial $Q_c$ can be found and subsequently improved through Algorithm 1. Also, to induce saturations on the states, we set the saturation bounds for the states to be $1$, $-0.1$, respectively.
The results on the state saturation errors are shown in Fig. \ref{fig11}. Also the effect of the proposed design on the control input and on the tracking performance are represented in Fig. \ref{fig12}. The impulse behavior in the control input is still observed but decreased when compared to the frequency-domain dynamic SANTW method. Apart from the peak, the control effort decreased in other times. On the other hand, the nominal reference tracking is improved. 

\textit{(d) Design of joint input-state H$_{\infty}$ anti-windup}

\begin{figure}[t]
\centering
\subfloat[]{\includegraphics[width=0.24\textwidth]{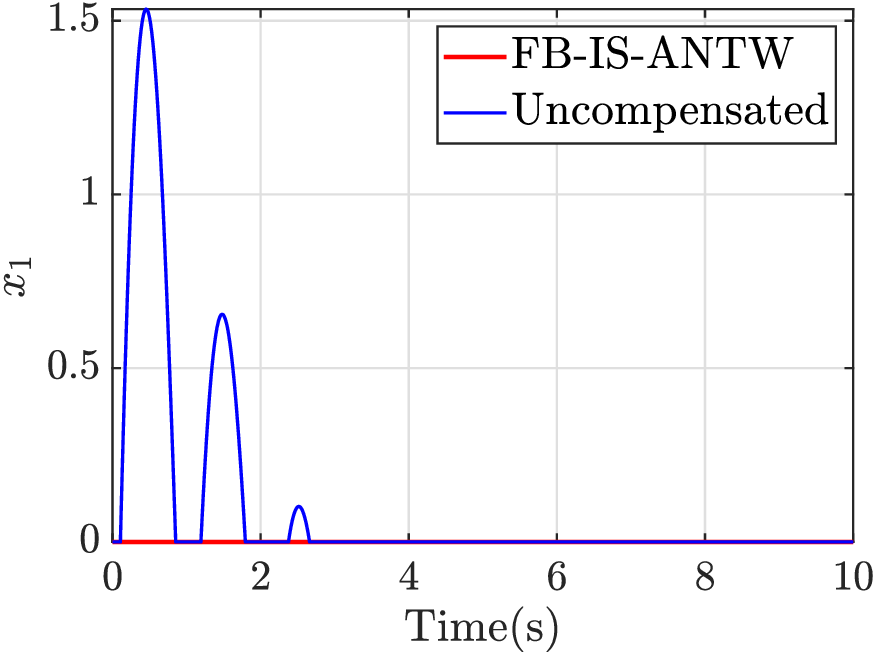}}
\subfloat[]{\includegraphics[width=0.24\textwidth]{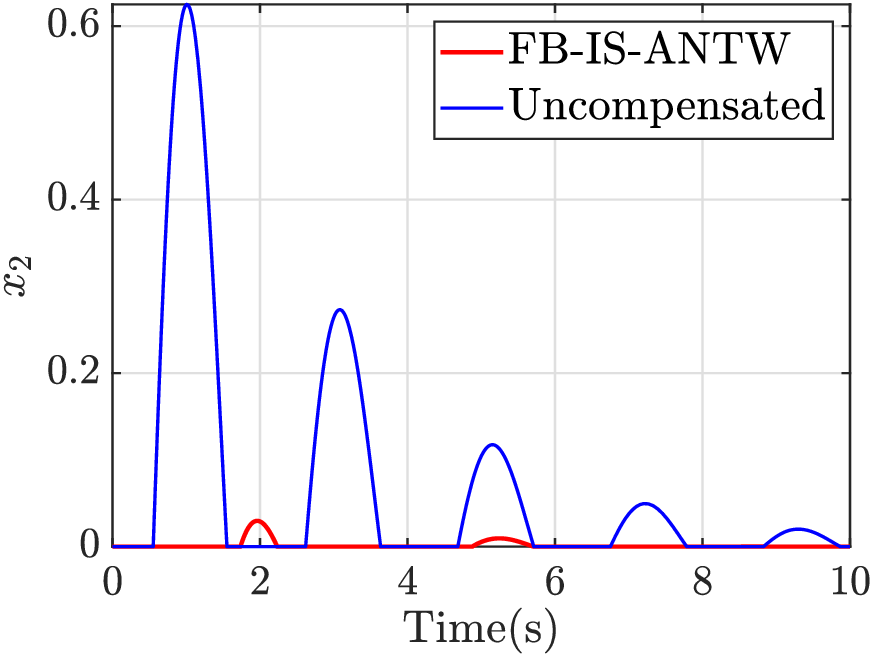}} 
\caption{Performance of full-block input-state anti-windup compensator to decrease the state saturation errors $\hat{x}-x$.}
\label{fig13}
\end{figure}

\begin{figure}[t]
\centering
\subfloat[]{\includegraphics[width=0.24\textwidth]{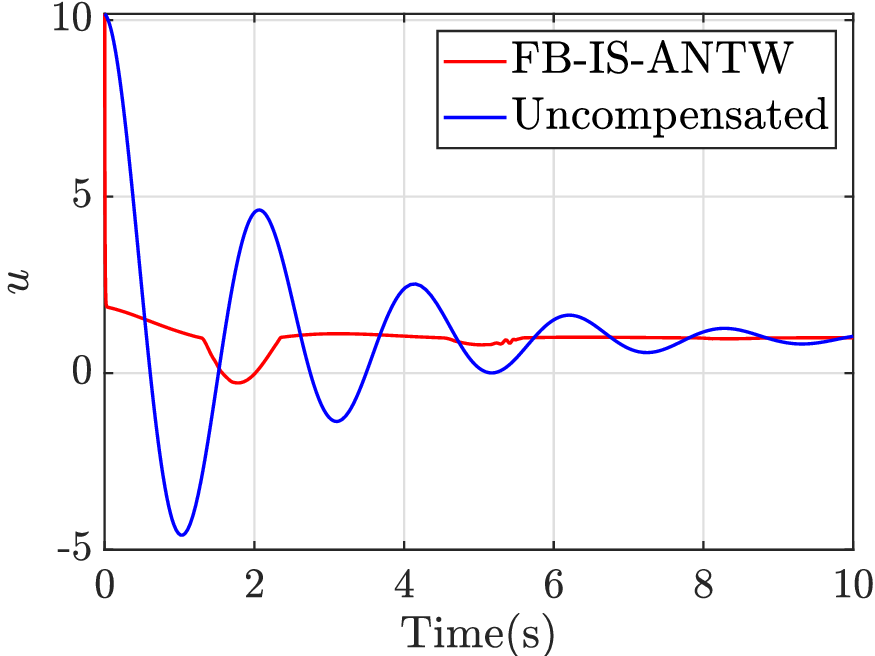}}
\subfloat[]{\includegraphics[width=0.24\textwidth]{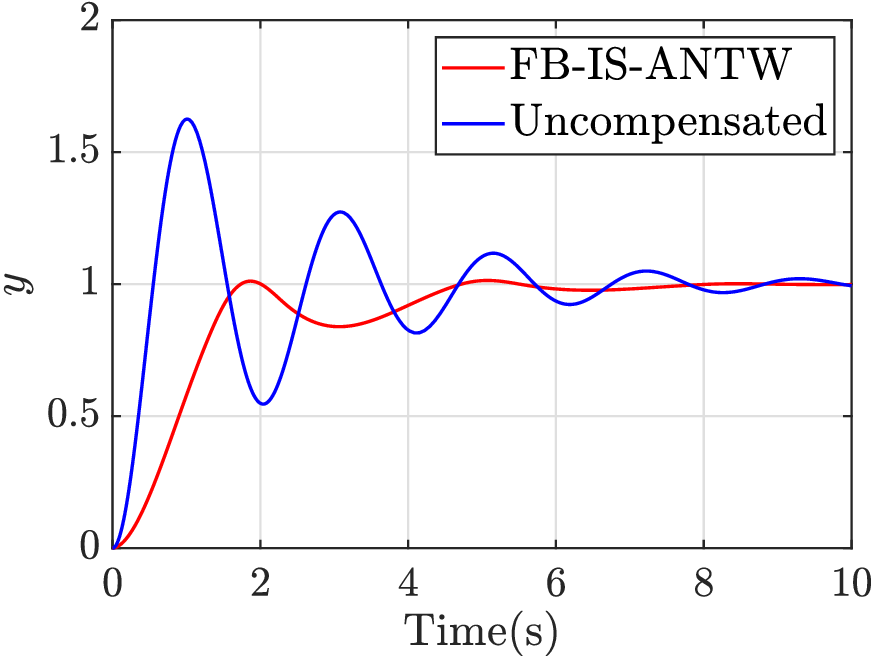}} 
\caption{Effects of full-block input-state anti-windup compensator on (a) control input and (b) closed-loop tracking.}
\label{fig14}
\end{figure}

In all previous cases, the controllers were designed based on minimizing the saturation error and, if possible, putting some penalty weights on the anti-windup intervention to avoid the potential input saturation. While, as mentioned in section 5, the IS-ANTW is designed by including the saturation block inside the controller which we referred to as soft-hard constrained compensation. Here, we first apply the method mentioned in \textit{Remark 6} using a full-matrix non-diagonal feedback. The weight functions are chosen as
\begin{equation}
   W_u=W_x=\frac{s+231.9}{s+22.74}
\label{eq24}
\end{equation}
The controllers (of the full-matrix feedback) are of order 6, attributed to its use of Riccatti-based frequency-domain $H_{\infty}$ optimization, is worth noting. The effectiveness of reducing the state saturation error $\hat{x}-x$ is demonstrated in Fig. \ref{fig13}, while the influence of IS-ANTW on the control input and reference tracking is depicted in Fig. \ref{fig14}(a) and (b), respectively. As seen, the control input remains limited and no spike behavior is observed. Also the tracking is improved. Despite the satisfactory performance of the compensator, its high order and complexity of implementation may not be desirable.

\begin{figure}[tb!]
\centering
\subfloat[]{\includegraphics[width=0.24\textwidth]{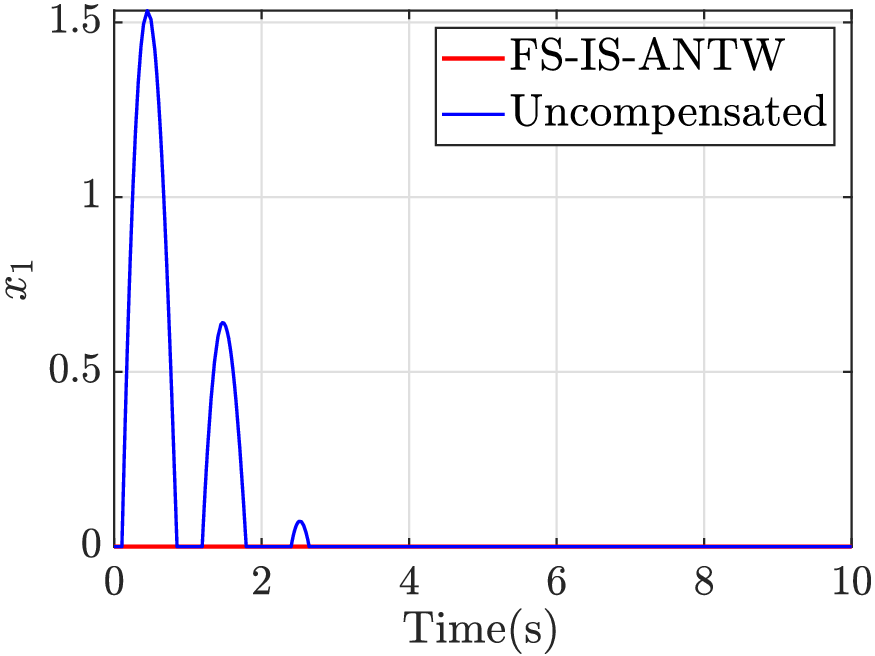}}
\subfloat[]{\includegraphics[width=0.24\textwidth]{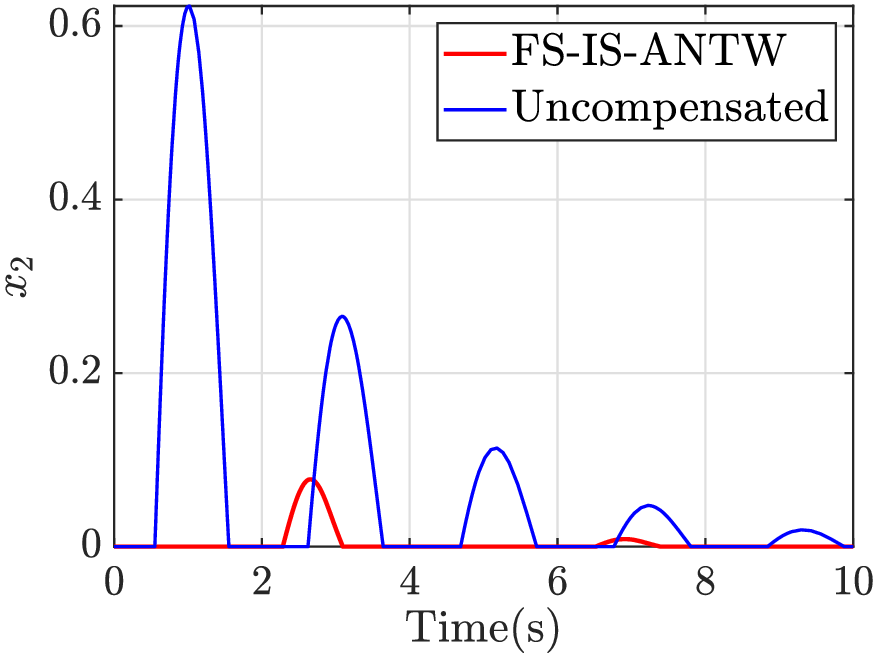}} 
\caption{Performance of fixed-structure input-state anti-windup to decrease the state saturation errors $\hat{x}-x$.}
\label{fig15}
\end{figure}

\begin{figure}[tb]
\centering
\subfloat[]{\includegraphics[width=0.24\textwidth]{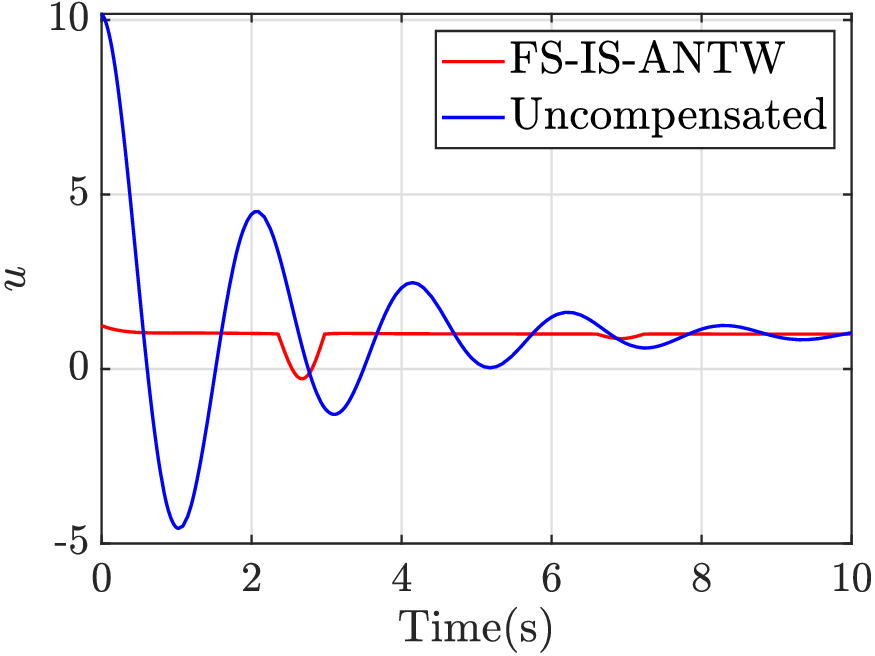}}
\subfloat[]{\includegraphics[width=0.24\textwidth]{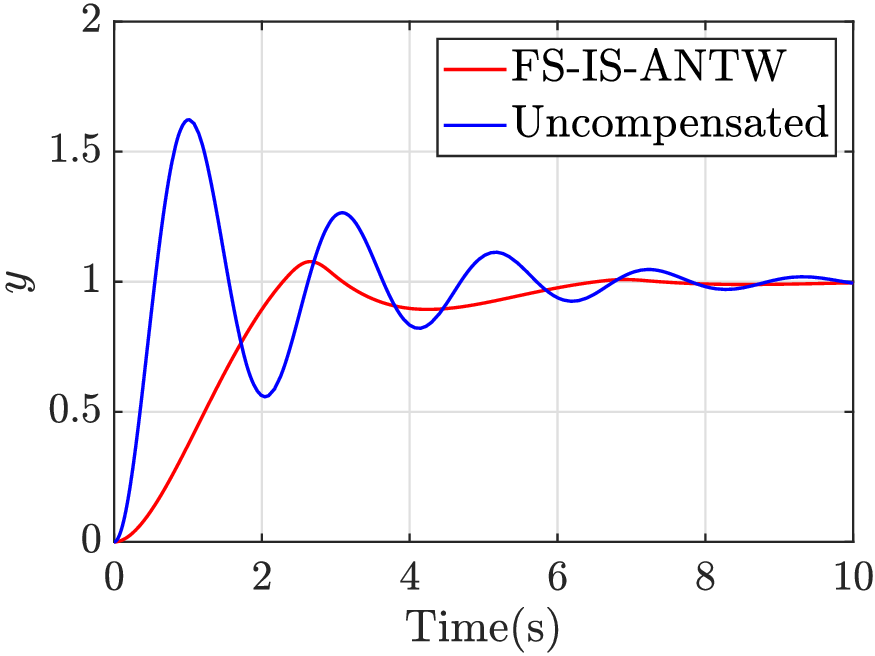}} 
\caption{Effects of fixed-structure input-state anti-windup on (a) control input and (b) closed-loop tracking.}
\label{fig16}
\end{figure}

As the last anti-windup controller designed, the IS-ANTW with a diagonal compensation structure (Fig. \ref{fig5}) is assessed. This structure ensures that the state/output saturation error is used independently for state anti-windup, and the input saturation error for input anti-windup. The controller is crafted using non-smooth optimization in MATLAB using \textit{hinfstruct} tool. Here, we selected an order of 6 for both the input compensator $G_{mu}$ and each of the state compensators $G_{mxi}$. The results showcasing the reduction in state saturation and the impact of the compensator on the overall control input and output tracking are displayed in Fig. \ref{fig15}, Fig. \ref{fig16}(a) and \ref{fig16}(b), respectively. It is evident that the saturation error is effectively mitigated, keeping the control input within acceptable limits, while also improving tracking. Despite a slight decline in performance (specifically in the saturation error of $x_2$) compared to full-matrix realizations, it offers advantages such as independent feedback, simple structure, and the capability of fixed-order synthesis, making it a preferred choice for dynamic systems with input and state constraints like power electronic inverters.

\begin{figure*}[t!]
\centering
\psfrag{P}[cc][cc][0.5][0]{$\textbf{PWM}$}
\psfrag{vdc}[cc][cc][1][0]{$V_{dc}$}
\psfrag{t}[cc][cc][1][0]{$\theta$}
\psfrag{PCC}[cc][cc][0.8][0]{$\text{PCC}$}
\psfrag{dq/abc}[cc][cc][0.9][0]{$dq/abc$}
\psfrag{abc/dq}[cc][cc][0.9][0]{$abc/dq$}
\psfrag{Controller}[cc][cc][1][0]{$\textbf{Controller}$}
\psfrag{SANTW}[cc][cc][0.85][0]{$\textbf{SANTW}$}
\psfrag{Grid}[cc][cc][1][0]{$\text{Grid}$}
\psfrag{vabc}[cc][cc][0.8][0]{$V_{abc}$}
\psfrag{vcabc}[cc][cc][0.8][0]{$\bm{\mathit{V_{c,abc}}}$}
\psfrag{iabc}[cc][cc][0.8][0]{$\bm{\mathit{i_{abc}}}$}    
\psfrag{igabc}[cc][cc][0.8][0]{$\bm{\mathit{i_{g,abc}}}$}
\psfrag{PLL}[cc][cc][0.75][0]{$\textbf{PLL}$}
\psfrag{LCL}[cc][cc][0.9][0]{$\text{LCL Filter}$}
\psfrag{C}[cc][cc][0.9][0]{$\text{DC-AC Inverter}$}
\psfrag{MS}[cc][lb][0.55][0]{$\textbf{Modulating signals}$}
\psfrag{TL}[cc][cc][0.9][0]{$\text{Transmission Line}$}
\includegraphics[width=0.7\textwidth]{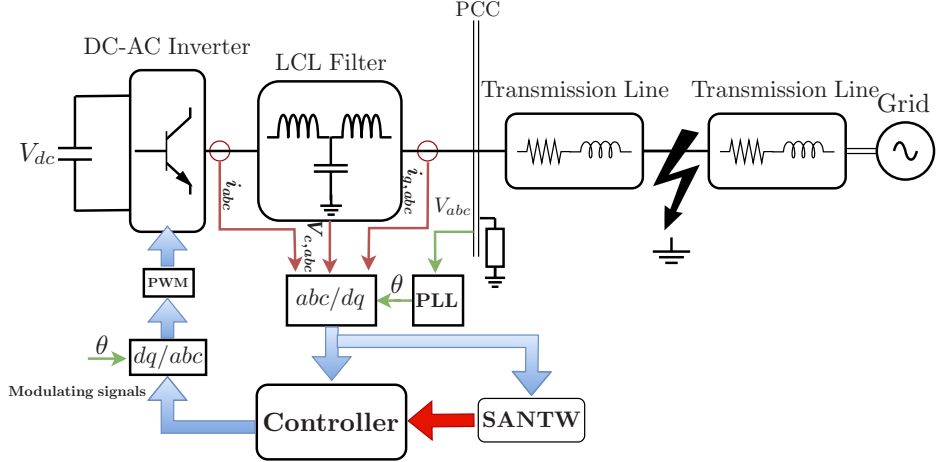}
\caption{Application of SANTW to limit the converter current in case of fault in the grid model.}
\label{fig17}
\end{figure*}

\subsection{Example II: Power control of grid-connected inverter-based resource (IBR) }
The concept of dynamic virtual power plant (DVPP), which is the center point of H2020 POSYTYF project, is described as an advanced system that integrates several decentralized power units, including renewable energy sources, storage systems, and flexible loads, to operate as a single entity \cite{marinescu2022dynamic}. Moreover, it dynamically adapts to real-time changes in energy demand and supply to enhance grid stability, efficiency, and reliability, while also providing essential ancillary services. Accordingly, the decentralized primary control of renewable energy resources plays an important role. Renewable sources are integrated to the AC grid through switching-type dc/ac converters including some RL or LCL filters to remove the high-frequency content. When the primary energy source is neglected, such a renewable source can be modeled by VSC grid-connected converter model given in 
Figure \ref{fig17}. It comprises a three-phase two-level VSC linked to the grid $V_{abc}$ via a transmission line (modeled as a R-L line equivalence). A variable load is connected to the Point of Common Coupling (PCC). The Phase Locked Loop (PLL) generates the grid voltage phase angle $\theta$, utilized for computing currents in the \textit{dq} frame ($i_{dq}$) from the three-phase currents ($i_{abc}$). Moreover, $\theta$ facilitates the conversion of the control signal from the \textit{dq} frame ($m_{dq}$) to the \textit{abc} frame ($m_{abc}$), thereby generating the modulating signals to regulate the converter. Additionally, an LCL filter is implemented to address the high-frequency oscillations in the output voltage of the IBR. 
The $dq$ state space equations are 
\begin{equation} \label{eq25}
\left\{ \begin{array}{l}
\frac{{d{i_{d1}}}}{{dt}} = -\frac{R_1}{L_1}{i_{d1}} + {\omega}_0 {i_{q1}} 
-\frac{v_{cfd}}{L_1}+\frac{1}{2}\frac{{{m_d}{V_{DC}}}}{L_1}\\
\frac{{d{i_{q1}}}}{{dt}} = -\frac{R_1}{L_1}{i_{q1}} - {\omega}_0 {i_{d1}}-\frac{v_{cfq}}{L_1}+\frac{1}{2}\frac{{{m_q}{V_{DC}}}}{L_1}\\
\frac{{d{i_{gd}}}}{{dt}} = -\frac{R_2}{L_2}{i_{gd}} + {\omega}_0 {i_{gq}} + \frac{v_{cfd}}{L_2}-\frac{1}{L_2}{v_{d}}\\
\frac{{d{i_{gq}}}}{{dt}} =-\frac{R_2}{L_2}{i_{gq}} - {\omega}_0 {i_{gd}} + \frac{v_{cfq}}{L_2}-\frac{1}{L_2}{v_{q}}\\
\frac{{d{v_{cfd}}}}{{dt}} = \frac{1}{C_f}{i_{d1}}-\frac{1}{C_f}{i_{gd}}+ {\omega}_{0}{v_{cfq}}\\
\frac{{d{v_{cfq}}}}{{dt}} = \frac{1}{C_f}{i_{q1}}-\frac{1}{C_f}{i_{gq}}- {\omega}_{0}{v_{cfd}}\\
\end{array} \right.
\end{equation}
and active and reactive powers are calculated as
\begin{equation}\label{eq26}
\begin{cases}
P = \frac{3}{2}({V_{d}}{i_{gd}} + {V_{q}}{i_{gq}}),\\
Q = \frac{3}{2}({V_{q}}{i_{gd}}-{V_{d}}{i_{gq}}).
\end{cases}
\end{equation}
\begin{figure}[htbp]
\centering
\subfloat[]{\includegraphics[width=0.24\textwidth]{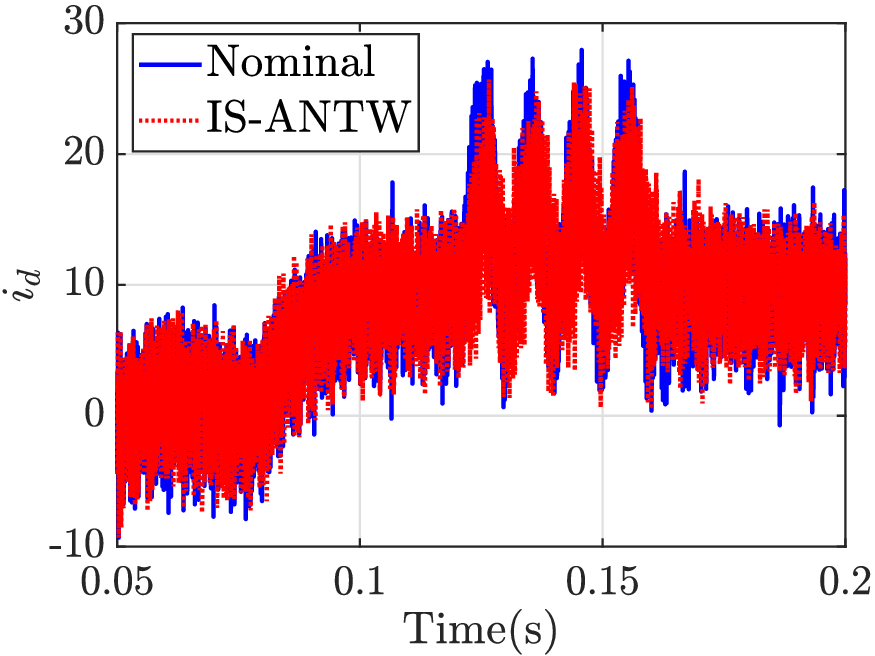}} 
\subfloat[]{\includegraphics[width=0.24\textwidth]{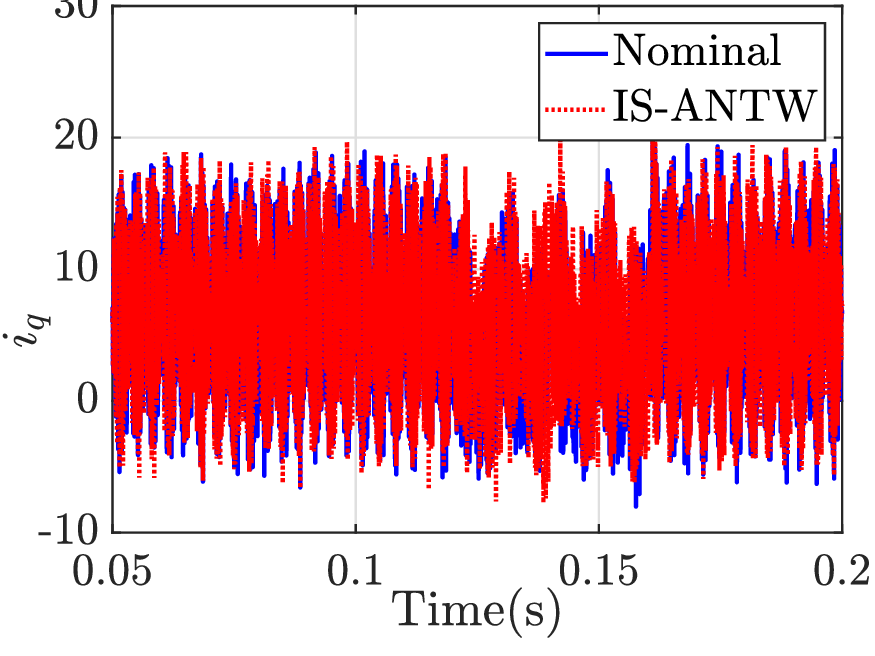}} \\
\subfloat[]{\includegraphics[width=0.24\textwidth]{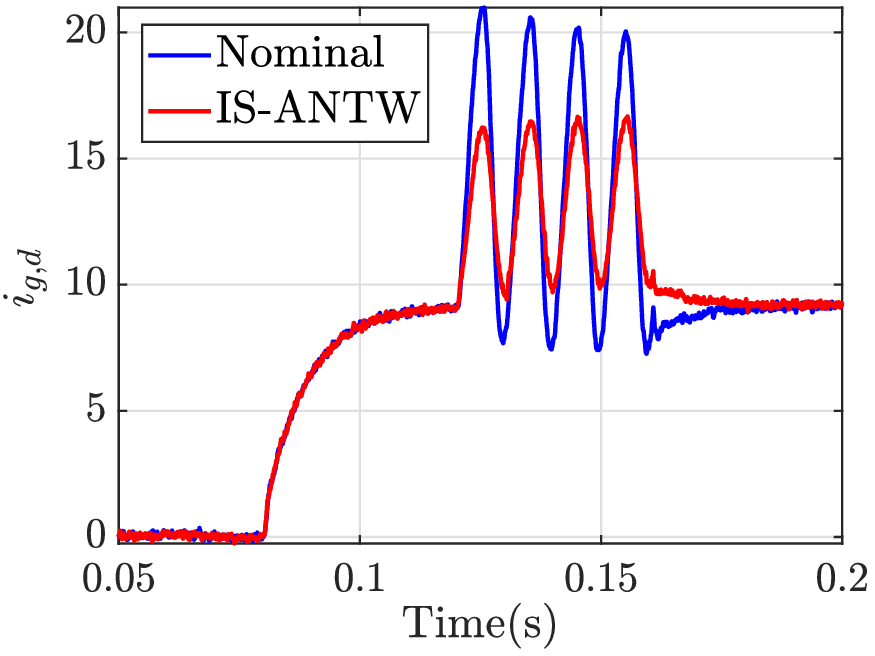}} 
\subfloat[]{\includegraphics[width=0.24\textwidth]{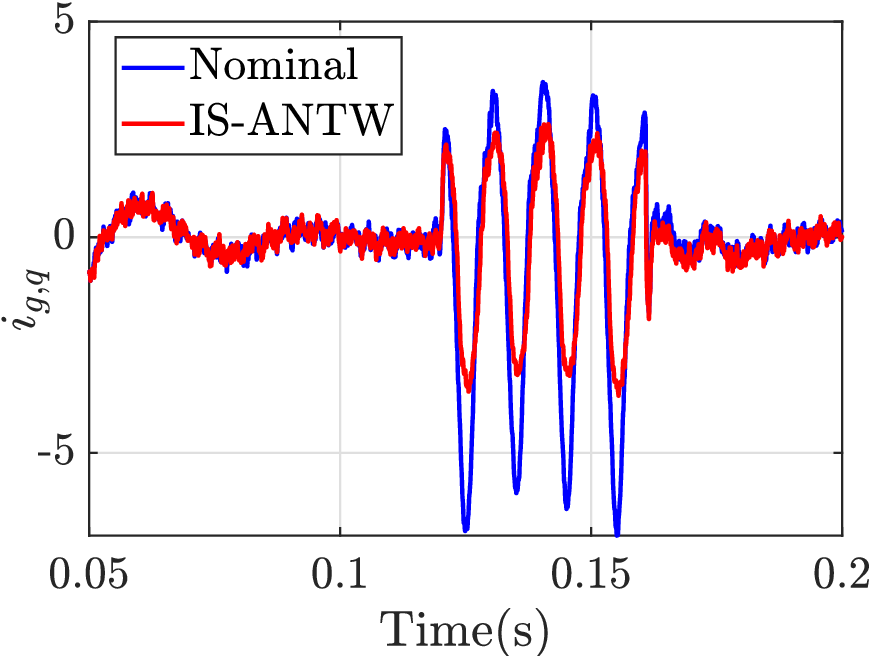}} \\
\subfloat[]{\includegraphics[width=0.24\textwidth]{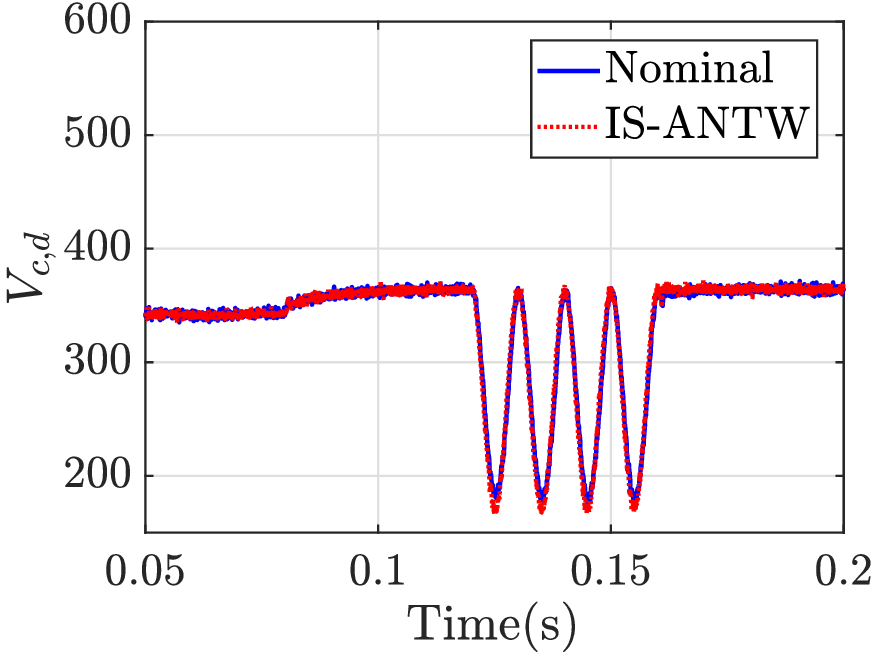}}
\subfloat[]{\includegraphics[width=0.24\textwidth]{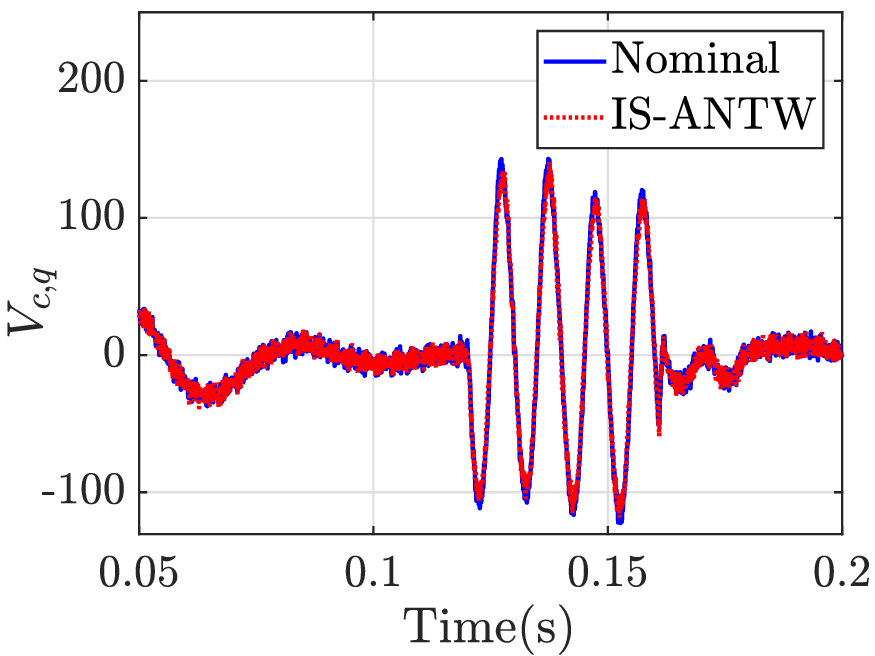}}
\caption{Performance of the dynamic state anti-windup compensator of fixed structure type in mitigiating the over-current of the grid-connected VSC as a result of the grid fault: (a),(b) inverter-side currents (c), (d) grid-side currents and (e), (f) filter capacitor volltage.}
\label{fig18}
\end{figure}
\begin{figure}[t]
    \centering
    \includegraphics[width=0.35\textwidth]{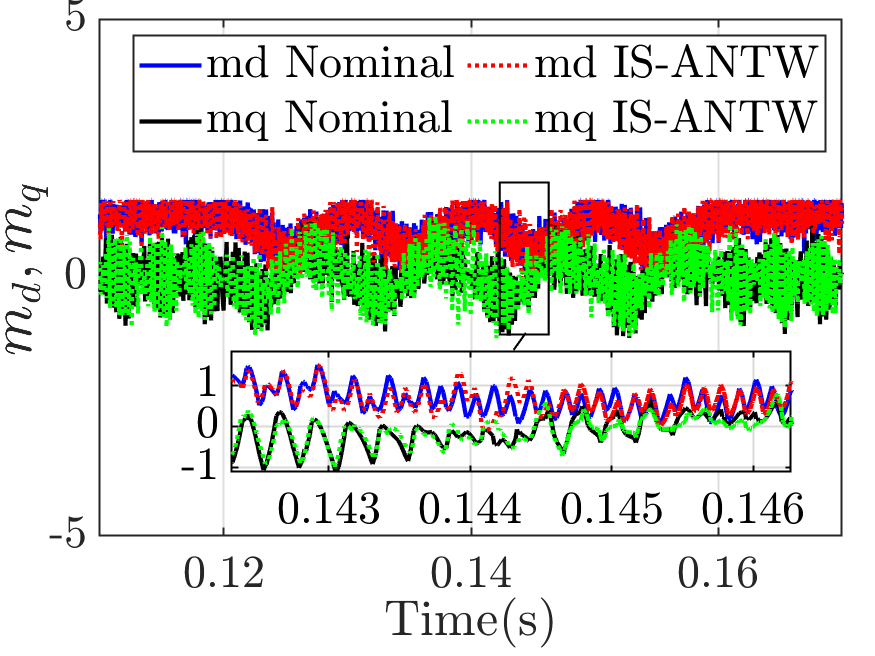}
    \caption{Comparison of modulating signals for the nominal and SANTW-compensated controllers}
    \label{fig19}
\end{figure} 
\begin{figure}[tb]
\centering
\subfloat[]{\includegraphics[width=0.24\textwidth]{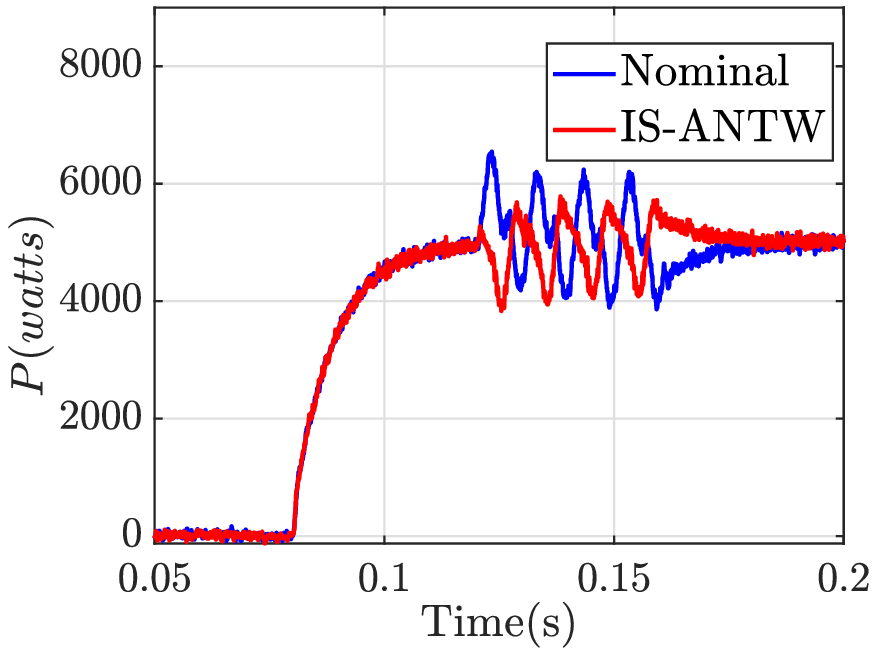}}
\subfloat[]{\includegraphics[width=0.24\textwidth]{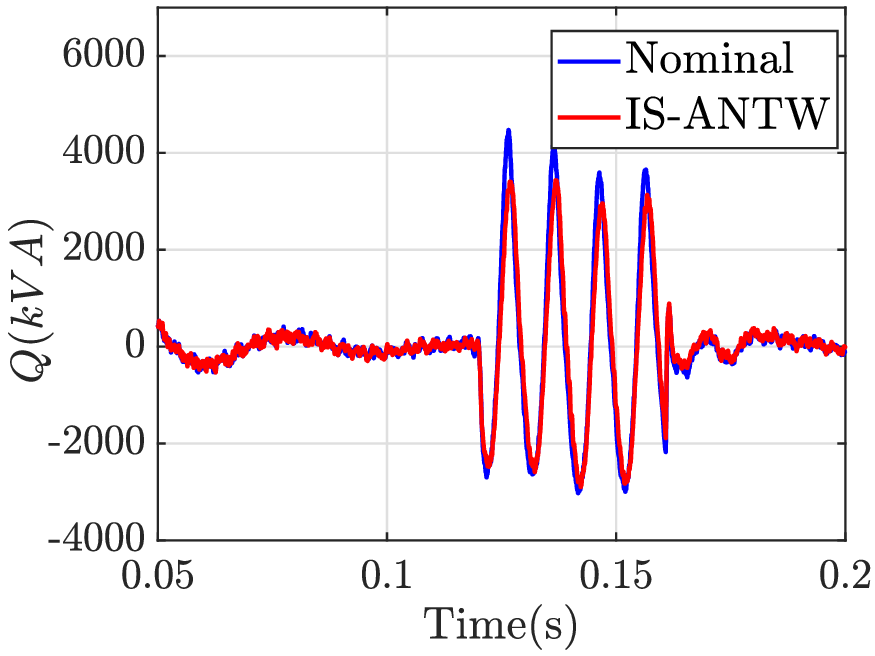}} 
\caption{Comparison of active and reactive power tracking for nominal and the one compensated by the fixed structure state anti-windup (a) power tracking (b) reactive power tracking}
\label{fig20}
\end{figure}
The control objective is to have $P{\rightarrow}P_{ref}$, $Q{\rightarrow}Q_{ref}$ in nominal operating condition and a good behaviour in case of severe grid disturbances like the short-circuits. On the other hand, there are limitations on PWM signals (control inputs) which are being applied to the switch gates, which should usually be between -1, 1.

To address this issue, the control strategy should incorporate a state anti-windup mechanism to manage current constraints. This SANTW compensation should be designed to maintain power tracking in nominal operation and limit degradation in case of grid faults.\\
To evaluate application of the proposed SANTW plans we design two scenarios: 

First we apply a step change at $t=0.08s$ in reference power to verify that the proposed SANTW compensator does not negatively affect the nominal power tracking. Indeed, as seen in Fig. 21, this brings the IBR from 0 to a nominal level of power. It is a way to examine if the integration of the SANTW compensator would downgrade the nominal power performance by the nominal controller. 
Second, while a constant nominal power is being injected into the grid, we introduce a short-circuit fault in the nearby grid to induce over-currents and assess how the SANTW can reduce these over-currents.

The fault applied is a phase-to-ground short-circuit implemented on the line connecting the converter to the ideal grid. This fault leads to an increase in the current drawn from the inverter, posing a potential threat to its operation. It may trigger protective measures, leading to the disconnection of the IBR from the grid. Consequently, this diminishes the resilience and fault-ride-through capability of the IBR, which are crucial for ensuring participation of renewable resources in grid ancillary services in contemporary power systems \cite{mohammed2023grid}. For more details on the control design for grid connected inverters see \cite{ngo2022emt, xing2022vsc, guo2018advanced} and the references therein. 

The fault is characterized by a resistance of $0.1$ ohms and a duration of $40$ milliseconds, occurring from $t=0.12s$ to $t=0.16s$. This fault results in oscillations in the system states. 
In \cite{ngo2022emt}, a $H_{\infty}$ controller is designed for (\ref{eq25}). When designing such a controller, frequency weights may be added on the output to achieve the desired frequency behavior of the loop transfer function. To modify the designed controller in \cite{ngo2022emt} to a state-constraint controller, we integrate the FS-IS-ANTW approach. We impose limitations only on the grid-side currents.

The results are shown in Fig. \ref{fig20}. Grid-side currents decrease due to IS-ANTW compensation. It should be noted that if there are protection limits for the current, these limits may be violated by the nominal controller but can be respected by the proposed anti-windup method. In this situation, the proposed FS-IS-ANTW prevents the renewable energy resource from disconnecting from the grid. Modulating signals, or control inputs, are depicted in Fig. \ref{fig19}. Adding input anti-windup compensation to the state anti-windup compensator keeps modulating signals within acceptable bounds. Fig. \ref{fig20} illustrates the impact of FS-IS-ANTW on power tracking. The proposed FS-IS-ANTW improves power tracking, reducing power oscillation during fault. 

\section{Conclusions}
The paper proposes a novel methodology called State Anti-Windup (SANTW) to address state constraints. By integrating saturation action within the compensator, SANTW effectively handles soft-hard constraints on states or outputs. The problem is formulated in a disturbance rejection framework, optimizing the energy transfer from the saturated state/output and nominal control to the saturation error and anti-windup action. \textcolor{black}{This provides a direct way to minimize the saturation. It is shown that a mixed input and state anti-windup is necessary to ensure performance. For that, a joint input-output anti-windup structure called IS-ANTW has been proposed along with methodologies to compute its gains.} 

Both frequency-domain 
$H_{\infty}$ synthesis and LMI-based synthesis solve the problem for static and dynamic compensators.
Although the $H_{\infty}$ framework was chosen in this paper, \textcolor{black} {the proposed IS-ANTW framework and concepts are not limited to this solution and can be addressed using several other methods. For example, existing advanced anti-windup techniques, such as Model Recovery Anti-Windup or co-prime factorization techniques or, moreover, approaches (mainly based on control barrier functions) mentioned in the introduction for soft state constraints can be used.}

Simulation results validate the effectiveness of all designed compensators. \textcolor{black} {The proposed methodology has also been used in an industrial application in the H2020 POSYTYF project. The proposed compensator has been applied to mitigate overcurrent in an inverter-based renewable energy resource during a nearby grid fault. The outcomes demonstrate the efficacy of FS-IS-ANTW in reducing the grid-fault-induced overcurrent, enabling the system to maintain its connection to the grid under fault conditions. This is a major advantage in promoting renewable energy generators. }

\begin{ack}                               
  This project has received funding from the European Union’s Horizon 2020 research and innovation programme under grant agreement No. 883985 (POSYTYF -- POwering SYstem ﬂexibiliTY in the Future through RES, https://posytyf-h2020.eu/).\\  
\end{ack}

\bibliographystyle{abbrvnat}        
 \bibliography{autosam}           

\appendix
\section{Appendix A: Proof of Theorem 1}
\vspace{-0.2cm}
Consider a quadratic Lyapunov function $V=x^TPx$ for some positive definite matrix $P$. Sufficient condition for the internal stability of the anti-windup closed-loop system and for achieving finite $\mathcal{L}_2$ gains from disturbance inputs to the outputs is presented as:
\begin{align}
    \label{a1}
    \Delta &=\dot{V}\left( x(t) \right)+\alpha \left\| {{u}_{m}} \right\|_{2}^{2}+\beta \left\| \hat{x}-x \right\|_{2}^{2} \nonumber \\ 
    & -{{\gamma}_{1}}\left\| {\hat{x}} \right\|_{2}^{2}+{{\gamma }_{2}}\left\| {{u}_{c}} \right\|_{2}^{2}<0,
\end{align}
Inequality (\ref{a1}) can be further extended as
\begin{align}
\label{a2}
\Delta   
 & =2{{x}^{T}}P\left[ \left( A-B{{K}_{m}} \right)x+B{u_c}+B{{K}_{m}}\hat{x} \right] \nonumber \\ 
 & +\beta{{\left( -x+\hat{x} \right)}^{T}}\left( -x+\hat{x} \right)  \\ 
 & +\alpha{{\left( -x+\hat{x} \right)}^{T}}K_{m}^{T}{{K}_{m}}\left( -x+\hat{x} \right) \nonumber-{{\gamma}_1} {{{\hat{x}}}^{T}}\hat{x}-{{\gamma}_2} {{u_c}^{T}}{u_c},  
\end{align}
Then it is straightforward to show that the inequality (\ref{a2}) is equivalent to 

\begin{equation}
    \label{a3}
    \resizebox{0.7\hsize}{!}{$
    \begin{aligned}
    {{J}_{1}} & = \left[ \begin{matrix}
        J_{1,11} & PB & PB{{K}_{m}} & -\beta I & -\alpha K_{m}^{T} \\
        * & -{{\gamma}_1} I & 0 & 0 & 0  \\
        * & * & -{{\gamma}_2} I & \beta I & \alpha K_{m}^{T}  \\
        * & * & * & -\beta I & 0  \\
        * & * & * & * & -\alpha I  \\
    \end{matrix} \right] < 0, \\
    J_{1,11} & = P(A - B{{K}_{m}}) + {{\left( A - B{{K}_{m}} \right)}^{T}}P,
    \end{aligned}
    $}
\end{equation}

using Schur’s complement. By pre- and post-multiplying (\ref{a3}) by $diag(Q, I, Q, I, I)$, where $Q=P^{-1}$, putting $Y={K_m}Q$ and using the inequality 

\begin{equation}
    \label{a5}
    -\gamma {{Q}^{T}}Q<-\gamma \left( {{Q}^{T}}{{Q}_{c}}+Q_{c}^{T}Q-Q_{c}^{T}{{Q}_{c}} \right)
\end{equation}
which holds for some $Q_c>0$,  we get the  LMI in (\ref{eq12}).

\section{Appendix B: Proof of Theorem 2}         
The augmented dynamics of the (\ref{eq13}) can be represented as                                      
 \begin{align}
 \label{a9}
  & \dot{\tilde{x}}=\tilde{A}\tilde{x}+{{{\tilde{B}}}_{u}}{{u}_{nom}}+{{{\tilde{B}}}_{s}}\hat{x}, \nonumber \\ 
 & z=\tilde{C}\tilde{x}+{{{\tilde{D}}}_{s}}\hat{x},  
 \end{align}
 where
 \begin{align}
     & \tilde{x}\triangleq \left[ \begin{matrix}
   x  \\
   q  \\
\end{matrix} \right],\,\,\tilde{A}=\left[ \begin{matrix}
   A-B{{K}_{m1}} & B{{K}_{m2}}  \\
   -{{B}_{q}} & {{A}_{q}}  \\
\end{matrix} \right],\,{{{\tilde{B}}}_{u}}=\left[ \begin{matrix}
   B  \\
   0  \\
\end{matrix} \right], \nonumber \\
& {{{\tilde{B}}}_{s}}=\left[ \begin{matrix}
   B{{K}_{m1}}  \\
   {{B}_{q}}  \\
\end{matrix} \right], 
  \tilde{C}=\left[ \begin{matrix}
   -I & 0  \\
   -{{K}_{m1}} & {{K}_{m2}} \nonumber \\
\end{matrix} \right],\,{{{\tilde{D}}}_{s}}=\left[ \begin{matrix}
   I  \\
   {{K}_{m1}} \nonumber \\
\end{matrix} \right],  \nonumber 
 \end{align}
The optimization problem is defined as
\begin{align}
    \label{a10}
    \underset{{{A}_{q}},{{B}_{q}},\,{{K}_{m1}},\,{{K}_{m2}}}{\mathop{Min}}\,\frac{{{\left\| z \right\|}_{2}}}{{{\left\| w \right\|}_{2}}},\,\,\,\,\,\,z=\left[ \begin{matrix}
   {{z}_{1}}  \\
   {{z}_{2}}  \\
\end{matrix} \right],\,\,w=\left[ \begin{matrix}
   {{u}_{c}}  \\
   {\hat{x}}  \\
\end{matrix} \right],	
\end{align}
By defining the Lyapunov function and following the same steps as in the previous section, the inequality condition for the problem of stability and disturbance attenuation is obtained as:

\begin{equation}
    \label{a11}
    \resizebox{0.85\hsize}{!}{$
    \begin{aligned}
    {\Gamma}=\left[ \begin{matrix}
   \tilde{A}Q+Q{{{\tilde{A}}}^{T}} & \left[ \begin{matrix}
   B  \\
   0  \\
\end{matrix} \right] & \left[ \begin{matrix}
   B{{K}_{m1}}  \\
   {{B}_{q}}  \\
\end{matrix} \right] & -\beta \left[ \begin{matrix}
   {{Q}_{1}}  \\
   {{h}_{1}}{{Q}_{1}}  \\
\end{matrix} \right] & \alpha \left[ \begin{matrix}
   -K_{m1}^{T}  \\
   K_{m2}^{T}  \\
\end{matrix} \right]  \\
   * & -{{\gamma }_{1}}I & 0 & 0 & 0  \\
   * & * & -{{\gamma }_{2}}I & \beta I & \alpha K_{m1}^{T}  \\
   * & * & * & -\beta I & 0  \\
   * & * & * & * & -\alpha I  \\
\end{matrix} \right]<0,
    \end{aligned}
    $}
\end{equation}

Then, by defining the matrix $Q$ as
\begin{equation}
    \label{a12}
    Q=\left[ \begin{matrix}
   {{Q}_{1}} & {{h}_{1}}{{Q}_{1}}  \\
   {{h}_{1}}{{Q}_{1}} & {{h}_{2}}{{Q}_{1}}  \\
\end{matrix} \right],\,\,{{h}_{2}}>h_{1}^{2}>0,
\end{equation}
Pre- and post- multiplying (\ref{a11}) by $diag(I,I,Q,I)$,and  then by utilizing the inequality (\ref{a5}) and defining ${{Y}_{1}}={{K}_{m1}}{{Q}_{1}},\,\,{{Y}_{2}}={{K}_{m2}}{{Q}_{1}},\,\,{{Y}_{A}}={{A}_{q}}{{Q}_{1}},$, the LMI condition (\ref{eq14}) in Theorem 2 can be obtained.

\end{document}